\documentclass[10pt,jpb]{iopart}

\usepackage{graphicx}

\usepackage{color}
\usepackage{ulem}

\begin{document}

\title[]{Long-time joint spectra and entanglement of two photoelectrons originating
in interacting auto-ionization systems}

\author{Jan Pe\v{r}ina Jr.}
\address{Institute of Physics of
Academy of Science of the Czech Republic, Joint Laboratory of
Optics, 17. listopadu 50a, 772 07 Olomouc, Czech Republic}
\ead{jan.perina.jr@upol.cz} \vspace{10pt}
\author{Anton\'{i}n Luk\v{s}}
\address{Palack\'{y} University, RCPTM, Joint Laboratory of Optics,
17. listopadu 12, 771 46 Olomouc, Czech Republic}
\author{Wieslaw Leo\'nski}
\address{Quantum Optics and Engineering Division, Institute of
Physics, University of Zielona G\'ora, Prof.~Z.~Szafrana 4a,
65-516 Zielona G\'ora, Poland}

\begin{indented}
\item[]
\end{indented}

\begin{abstract} Two auto-ionization systems in a stationary optical field
mutually interacting via the dipole-dipole interaction are
considered. Their evolution is analytically found. Joint spectra
of two ionized electrons are analyzed in detail in the long-time
limit for comparable strengths of direct and indirect ionization
paths as well as the dominating indirect ionization path.
Entanglement in the state of two ionized electrons is quantified
using the density of quadratic negativity. Suitable conditions for
obtaining highly entangled states are discussed.
\end{abstract}

\pacs{33.80.Eh,34.20.-b,03.65.Ud,32.80.-t}


\vspace{2pc} \noindent{\it Keywords}: two-electron ionization
spectra, auto-ionization, dipole-dipole interaction, Fano model,
bipartite entanglement, quadratic negativity

\maketitle

\ioptwocol

\section{Introduction}

Ionization is one of the most interesting physical processes
arising from the interaction of atoms or molecules with
electromagnetic fields. Ionized electrons are characterized by
their ionization spectra that contain information about the
electronic structure of the ionized atoms or molecules. For this
reason, ionization processes have been used as a strong tool for
the investigation of electronic structures
\cite{Journel1993,Durand2001}. The ionization spectra are strongly
sensitive to individual bound excited electronic states through
which the ionization can efficiently proceed (for an extended list
of references, see e.g.
\cite{Agarwal1984,Leonski1987,Leonski1988,Leonski1988a}).

The mechanism of auto-ionization occurring with the contribution
of a bound excited state has been extensively studied for He atoms
containing only two electrons. Quantum mechanical calculations
have shown that there is a large probability of exciting both
electrons together \cite{Aberg1970} due to the strong
electron-electron correlations present in the electronic ground
state \cite{Byron1967}. Energy of one of the excited electrons
given to the other excited electron then allows ionization of this
electron. The first electron returns to its ground state. This
process modelled in \cite{Wickenhauser2005,Zhao2005,Tong2005} with
the inclusion of an additional probe field has recently been
observed using attosecond pump-probe spectroscopic techniques
\cite{Eckle2008,Gilbertson2010}.

The presence of an auto-ionizing state in atoms manifests itself
in the ionization spectra by the so-called Fano zero
\cite{Fano1961}. It denotes an electron energy missing in the
ionization spectrum. This occurs due to the completely destructive
interference between different paths of ionization, including the
interference between direct and indirect (through an auto-ionizing
state) ionization processes \cite{Lambropoulos1981}. This
interference also results in sharp peaks in the ionization spectra
caused by intense laser pulses
\cite{Rzazewski1981,Lewenstein1983}. As predicted by Fano
\cite{Fano1961}, $ n $ zeroes will be present for $ n $
auto-ionizing levels for an atom as long as only a single
ionization channel is available. They are replaced by local minima
when more ionization channels are present, and they may even
disappear in molecules because of the presence of nuclear motion
\cite{Palacios2013}.

The presence of auto-ionizing states influences considerably many
physical effects. For example, auto-ionization systems have been
investigated as media exhibiting electromagnetically-induced
transparency
\cite{Enk1994,Raczynski2006,Dinh2014}. 
They have also been applied for slowing down the light propagating
through the medium with auto-ionizing states \cite{Raczynski2006}.
Moreover, special attention has been devoted to the problem of
near-threshold ionization \cite{Raczynski1993} including quantum
anti-Zeno effect \cite{Lewenstein2000}. The role of elastic
collisions and finite laser spectral width has been analyzed for
such systems in \cite{Rzazewski1983}. Auto-ionization has even
been considered for weak quantum fields \cite{Leonski1990}, in
particular, for squeezed light \cite{Leonski1993}. Interestingly,
Fano profiles and Fano resonances can appear not only in typical
atomic/optical models. They have been found, for instance, in
plasmonic systems \cite{Ridolfo2010,Giannini2011}, various
nanoscale structures \cite{Miroshnichenko2010} including quantum
dots \cite{Karwacki2013} or in a broad range of superconducting
systems \cite{Fedotov2010,Fogelstrom2014,Szczesniak2014}. We note
that states of the continuum and bound states participate together
in the effect called the Feshbach resonance \cite{Feshbach1962}.
In this effect, a quasi-bound molecular state is formed from the
state of two free atoms. This results in considerable enhancement
of inelastic collision processes \cite{Inouye1998}.

Ionization can also be observed due to the electronic Coulomb
interaction between two excited electrons being at two neighbor
atoms. In this case, energy of one excited electron is given to
the excited electron on the neighbor atom at which ionization
occurs \cite{Cederbaum1997}. If two excited electrons emerge from
the interaction with the strong electromagnetic field, we have
two-center resonant photo-ionization
\cite{Najjari2010,Voitkiv2010}. Also two-center dielectronic
recombination \cite{Muller2010,Voitkiv2010} which is just the
inverse process in which a free electron is captured by a
positively charged system of two neighbor atoms has been
suggested. These effects are important in atomic and molecular
clusters in which the exponential character of these processes can
be modified by additional interactions, e.g. by capturing the
ionized electron \cite{Averbukh2010}.

In an intense external field, double ionization may occur leaving
two free ionized electrons \cite{Aberg1970}. It has been
extensively studied in isolated atoms or molecules
\cite{Walker1994}. The dynamics of double ionization is
predominantly governed by the strong Coulomb interaction among the
electrons and the core of an atom. As a consequence, strong
correlations in momenta of two ionized electrons occur
\cite{Weber2000,Rudenko2007}. Considering He atoms, rotational and
bending modes of the two-excited-electron motion preceded the
ionization have been identified \cite{Morishita2007}. Double
ionization of Ne atoms has been experimentally studied in
\cite{Uiberacker2007}.

Here, we consider another process in which two electrons are
ionized in such a way that their energies are highly correlated.
Contrary to the double ionization model of a single atom, we
consider a system of two neighbor atoms or molecules under the
interaction with a strong cw external field. Each atom or molecule
provides one ionized electron. We assume here that each atom or
molecule can be ionized through an auto-ionizing level or by the
process of direct ionization. We show that, as a result of the
dipole-dipole interaction between two atoms or molecules, the
state of two ionized electrons can be strongly entangled and both
electrons can exhibit strong spectral correlations.

The model considered here is analyzed using the algebraic method
applied already by Fano \cite{Fano1961}. Our model represents a
generalization of the previously developed models in which
ionization spectra and correlations between an ionized electron
and a bound electron found at a neighbor atom have been studied.
It has been shown that the so-called dynamical Fano zeroes
occurring once per the Rabi period can be observed in conditional
ionization spectra, independently on the presence
\cite{PerinaJr2011a} or absence \cite{PerinaJr2011} of an
auto-ionizing level. Moreover, for the previously discussed
systems, the dipole-dipole interaction leads to the generation of
entangled states encompassing both the ionized and the bound
electrons \cite{Luks2012,Perinova2014}. The observed spectral
correlations have been conveniently quantified by the density of
quadratic negativity \cite{Hill1997} defined in terms of partially
transposed statistical operators \cite{Peres1996,Horodecki1996}.
Moreover, it has been shown that the dipole-dipole interaction
does not suppress the occurrence of the Fano zero in general, but
its existence is restricted to only rather special conditions
\cite{PerinaJr2011c,PerinaJr2011b}. Also a quantized field has
been considered for the two-atom auto-ionizing model showing that
many sharp peaks in the ionization spectra originating in discrete
field energies can occur \cite{Perinova2014}.

It will be shown that the exact Fano zeroes typical for ionization
spectra of individual atoms cannot occur in the analyzed model of
two auto-ionization systems. On the other hand, many spectral
features found in the Fano model of an auto-ionization system
\cite{Fano1961} are present in the analyzed system including sharp
peaks and dips. The applied algebraic approach provides analytical
formulas that allow to get a deep insight into the behavior of two
interacting auto-ionization subsystems analyzing spectral
correlations and entanglement \cite{Ficek2010} of the
state of two ionized electrons.  
Similarly as in \cite{Luks2012} the density of quadratic
negativity is applied to quantify quantum spectral correlations.
Although the model is discussed in general using the values of
parameters emphasizing its main features, the values of parameters
suitable for molecular condensates (including molecular crystals)
\cite{Silinsh1994} are also considered.

The paper is organized as follows. Hamiltonian of two coupled
auto-ionization systems in a stationary optical field is given in
Sec.~2, together with the corresponding dynamical equations and
their solution. Quadratic negativity and its density as
quantifiers of entanglement between two ionized electrons is
introduced in Sec.~3. Long-time photoelectron ionization spectra
for the cases of comparable direct and indirect ionization paths
(Subsec.~4.1) and dominating indirect ionization path
(Subsec.~4.2) are analyzed in Sec.~4. Entanglement in the
long-time photoelectron spectra is discussed in Sec.~5. Sec.~6 is
devoted to the process of ionization in molecular condensates.
Sec.~7 brings conclusions. Analytical solution of the model
without the mutual interaction is provided in Appendix~A.
Appendix~B is devoted to the competition of the dipole-dipole
interactions between discrete auto-ionizing levels and with the
continua.

\section{Hamiltonian, dynamical equations and their solution}

\begin{figure}  
 \centerline{\includegraphics[scale=0.6]{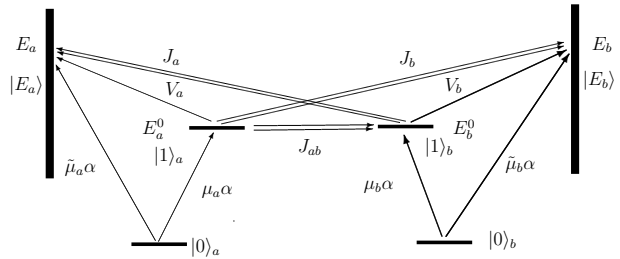}}
 \caption{Scheme of the mutually interacting auto-ionization systems $ a $ and $ b $.
  State $ |1\rangle_j $ denotes an excited bound state
  of atom $ j $ with energy $ E_j $ whereas state $ |E_j\rangle $
  with energy $ E_j $ lies inside the continuum of atom $ j $, $j=a,b$.
  Symbols $ \mu_j $ and $ \tilde\mu_j $ stand for the dipole moments
  between the ground state $ |0\rangle_j $ and the corresponding excited states, $ \alpha_L $ stands for
  the pumping amplitude,
  $ V_j $ describes the Coulomb configurational coupling
  between states $ |1\rangle_j $ and $ |E_j\rangle $. Symbols
  $ J_{ab} $, $ J_a $ and $ J_b $ refer to the dipole-dipole interactions
  between the excited/ionized states at atoms $ a $ and $ b $
  indicated in the scheme. Double arrows mean that two electrons at atoms
  $ a $ and $ b $ participate in the interaction leading to energy transfer.}
\label{fig1}
\end{figure}

We consider two auto-ionization systems describing, e.g., two
atoms or molecules that mutually interact by the dipole-dipole
interaction (for the scheme, see Fig.~\ref{fig1}).
Both the ionization systems denoted as $ a $ and $ b $ are assumed
to have one auto-ionizing level. They interact with a stationary
optical field with amplitude $ \alpha_L $ and frequency $ E_L $
through the corresponding dipole moments. Hamiltonian $
\hat{H}_{\rm j} $ describing auto-ionization system $ j $, $ j=a,b
$, can be written as ($ \hbar = 1 $ is assumed,
\cite{Meystre2007}):
\begin{eqnarray}   
 \hat{H}_{\rm j} &=& \hat{H}_{\rm j}^0 + \hat{H}_{\rm j}^L , \hspace{5mm} j=a,b,
\label{1}  \\
 \hat{H}_{\rm j}^0 &=& E_j^0|1\rangle_j{}_j\langle1| +
  \int dE_j  E_j|E_j\rangle \langle E_j| \nonumber \\
 & & \mbox{} + \int dE_j \left[ V_j |E_j\rangle\,{}_j\langle 1|+\mbox{H.c.}
  \right] ,
\label{2}  \\
 \hat{H}_{\rm j}^L &=& \Bigl[ \mu_j \alpha_L\exp(-iE_L t) |1\rangle_j{}_j\langle 0| \nonumber \\
 & & \hspace{-4mm} \mbox{} +
  \int dE_j \tilde{\mu}_j \alpha_L\exp(-iE_L t) |E_j\rangle\,{}_j\langle 0|+\mbox{H.c.}
  \Bigr] .
\label{3}
\end{eqnarray}
In Eqs.~(\ref{1}---\ref{3}), Hamiltonian $ \hat{H}_{\rm j}^0 $
describes the inner structure of auto-ionization system $ j $
whereas Hamiltonian $ \hat{H}_{\rm j}^L $ arises from the
interaction of the system with the optical field. Symbol $ E_j^0 $
denotes the excitation energy from the ground state $ |0\rangle_j
$ into the excited bound state $ |1\rangle_j $ of atom $ j $. The
continuum of atom $ j $ is formed by states $ |E_j\rangle $ having
energies $ E_j $. The Coulomb configuration interaction between
states $ |1\rangle_j $ and $ |E_j\rangle $ inside the continuum is
described by coupling constant $ V_j $. Dipole moments $ \mu_j $
and $ \tilde\mu_j $ characterize an optical excitation of the
corresponding states. Symbol $ \mbox{H.c.} $ stands for the
Hermitian conjugated term.

The dipole-dipole interaction between atoms $ a $ and $ b $ leads
to energy transfer \cite{Silinsh1994}. It can be described by the
following Hamiltonian $ \hat{H}_{\rm trans} $:
\begin{eqnarray}   
 \hat{H}_{\rm trans} &=& \Bigl[ J_{ab} |0\rangle_a |1\rangle_b {}_b\langle0|{}_a\langle1|
  \nonumber \\
 & & \mbox{} + \int dE_a J_a |E_a\rangle\, |0\rangle_b  {}_b\langle1| {}_a\langle0|
  \nonumber \\
 & & \mbox{} + \int dE_b J_b |0\rangle_a |E_b\rangle\, {}_b\langle0|{}_a\langle1|
  + \mbox{H.c.} \Bigr] .
\label{4}
\end{eqnarray}
In this interaction, one electron loses its energy and returns
into its ground state, whereas the other electron takes this
energy and moves into its own excited/ionized state. Constant $
J_{ab} $ describes the interaction between excited states $
|1\rangle_a $ and $ |1\rangle_b $ whereas constant $ J_a $ [$ J_b
$] characterizes the interaction between excited state $
|1\rangle_b $ [$ |1\rangle_a $] and ionized states $ |E_a\rangle $
[$ |E_b\rangle $]. We assume that the electrons in ionized states
do not mutually interact.

A general quantum state of two electrons at atoms $ a $ and $ b $
written in the frame rotating with frequency $ E_L $ can be
expressed as:
\begin{eqnarray}   
 |\psi\rangle(t) &=& c_{00}(t) |0\rangle_{a} |0\rangle_{b}
   + c_{10}(t) |1\rangle_{a} |0\rangle_{b} \nonumber \\
 & & \hspace{0mm} \mbox{} + c_{01}(t) |0\rangle_{a} |1\rangle_{b}
   + c_{11}(t) |1\rangle_{a} |1\rangle_{b} \nonumber \\
 & & \hspace{-7mm} \mbox{}  + \int dE_a \left[ d_{a,0}(E_a,t)
  |E_a\rangle |0\rangle_b + d_{a,1}(E_a,t)
  |E_a\rangle |1\rangle_b \right] \nonumber \\
 & & \hspace{-7mm} \mbox{}  + \int dE_b \left[ d_{b,0}(E_b,t)
  |0\rangle_a |E_b\rangle  + d_{b,1}(E_b,t)
  |1\rangle_a |E_b\rangle \right] \nonumber \\
 & & \hspace{0mm} \mbox{}  + \int dE_a \int dE_b \, d(E_a,E_b,t)
  |E_a\rangle |E_b\rangle.
\label{5}
\end{eqnarray}
Time-dependent coefficients $ c_{00} $, $ c_{10} $, $ c_{01} $, $
c_{11} $, $ d_{a,0}(E_a) $, $ d_{a,1}(E_a) $, $ d_{b,0}(E_b) $, $
d_{b,1}(E_b) $, and $ d(E_a,E_b) $ characterize the state $
|\psi\rangle $ at an arbitrary time $ t $. They can be
conveniently grouped into the vectors $ {\bf c}^T(t) = [c_{00}(t),
c_{10}(t), c_{01}(t), c_{11}(t)] $ and $ {\bf d}_j^T(E_j,t) = [
d_{j,0}(E_j,t), d_{j,1}(E_j,t)] $ for $ j=a,b $. Symbol $ T $
denotes transposition. Normalization of the state $
|\psi\rangle(t) $ in Eq.~(\ref{5}) provides the relation $ {\bf
c}^\dagger(t) {\bf c}(t) + \sum_{j=a,b} \int dE_j {\bf
d}_j^\dagger (E_j,t) {\bf d}_j(E_j,t) + \int dE_a \int dE_b
|d(E_a,E_b,t)|^2 = 1 $.

The Schr\"odinger equation with the overall Hamiltonian $
\hat{H}_a + \hat{H}_b + \hat{H}_{\rm trans} $ provides the
following dynamical equations for the coefficients characterizing
state $ |\psi\rangle(t) $ written in Eq.~(\ref{5}):
\begin{eqnarray} 
 i \frac{d}{dt} \left[
  \begin{array}{c} {\bf c}(t) \cr {\bf d}_a(E_a,t) \cr
   {\bf d}_b(E_b,t) \cr d(E_a,E_b,t) \end{array} \right]
   &=& \left[ \begin{array}{cc}
    {\bf A} & \int dE_a {\bf B}_a  \cr
    {\bf B}_a^\dagger &  (E_a-E_L) {\bf 1} + {\bf K}_b  \cr
    {\bf B}_b^\dagger & {\bf 0} \cr
    {\bf 0} & {\bf I}_b^\dagger \end{array}
     \right.  \nonumber \\
  & & \hspace{-39mm}  \left. \begin{array}{cc}
    \int dE_b {\bf B}_b & {\bf 0} \cr
    {\bf 0} & \int dE_b {\bf I}_b \cr
    (E_b-E_L) {\bf 1} + {\bf K}_a & \int dE_a {\bf I}_a \cr
    {\bf I}_a^\dagger & E_a+E_b-2E_L \end{array} \right]
    \left[ \begin{array}{c} {\bf c}(t) \cr {\bf d}_a(E_a,t) \cr
     {\bf d}_b(E_b,t) \cr d(E_a,E_b,t) \end{array} \right].
     \nonumber \\
  & &
\label{6}
\end{eqnarray}
In Eq.~(\ref{6}), symbol $ {\bf 1} $ ($ {\bf 0} $) stands for the
unity (zero) matrix of appropriate dimension(s).

Matrices $ {\bf A} $, $ {\bf B}_a $, $ {\bf B}_b $, $ {\bf K}_a $,
$ {\bf K}_b $, $ {\bf I}_a $ and $ {\bf I}_b $ introduced in
Eq.~(\ref{6}) are defined as follows:
\begin{eqnarray}  
& & {\bf A} = \left[\begin{array}{cccc}
  0 & \mu_a^*\alpha_L^* & \mu_b^*\alpha_L^* & 0 \\
  \mu_a\alpha_L & \Delta E_a^0 & J_{ab}^* & \mu_b^*\alpha_L^* \\
  \mu_b\alpha_L & J_{ab} & \Delta E_b^0 & \mu_a^*\alpha_L^* \\
  0 & \mu_b\alpha_L & \mu_a\alpha_L & \Delta E_a^0 + \Delta E_b^0
  \end{array}\right] , \nonumber \\
 & & {\bf B}_a = \left[\begin{array}{cc}
   \tilde\mu_a^*\alpha_L^* & 0 \\   V_a^* & 0 \\
   J_a^* & \tilde\mu_a^*\alpha_L^* \\
   0 & V_a^* \end{array}\right], \hspace{3mm}
    {\bf B}_b = \left[\begin{array}{cc}
     \tilde\mu_b^*\alpha_L^* & 0 \\ J_b^* & \tilde\mu_b^*\alpha_L^* \\
     V_b^* & 0 \\ 0 & V_b^* \end{array}\right], \nonumber \\
 & & {\bf K}_j = \left[\begin{array}{cc} 0 & \mu_j^*\alpha_L^* \\
   \mu_j\alpha_L & \Delta E_j^0 \end{array}\right] , \hspace{3mm}
  {\bf I}_j = \left[\begin{array}{c} \tilde\mu_j^*\alpha_L^* \\
   V_j^* \end{array}\right], \nonumber \\
 & &  \hspace{25mm} j=a,b;
\label{7}
\end{eqnarray}
$ \Delta E_j^0 = E_j^0 - E_L $, $ j=a,b $.

The set of equations (\ref{6}) can be conveniently solved by the
Laplace transform defined for an arbitrary function $ f(t) $ as $
\tilde f(\varepsilon) = \int_{0}^{\infty} dt f(t)
\exp(i\varepsilon t) $. If we assume that only the discrete levels
of atoms $ a $ and $ b $ are initially populated, we arrive at the
following set of equations:
\begin{eqnarray} 
 \left[\varepsilon {\bf 1} - {\bf A} \right] \tilde{\bf c}(\varepsilon)
   - \sum_{j=a,b} \int dE_j {\bf B}_j \tilde{\bf d}_j(E_j,\varepsilon) = i{\bf c}(0) ,
\label{8}  \\
 \left[ (\varepsilon - E_a + E_L) {\bf 1} - {\bf K}_b \right]
  \tilde{\bf d}_a(E_a,\varepsilon) \hspace{0mm}  \nonumber \\
 \mbox{}  \hspace{10mm} - {\bf B}_a^\dagger \tilde{\bf c}(\varepsilon)
   - \int dE_b {\bf I}_b \tilde{d}(E_a,E_b,\varepsilon) = {\bf 0} ,
\label{9} \\
 \left[ (\varepsilon - E_b + E_L) {\bf 1} -
  {\bf K}_a \right] \tilde{\bf d}_b(E_b,\varepsilon) \hspace{0mm} \nonumber \\
 \mbox{} \hspace{10mm} - {\bf B}_b^\dagger \tilde{\bf c}(\varepsilon)
   - \int dE_a {\bf I}_a \tilde{d}(E_a,E_b,\varepsilon) = {\bf 0} ,
\label{10} \\
 (\varepsilon - E_a - E_b + 2E_L) \tilde{d}(E_a,E_b,\varepsilon)
   \hspace{0mm} \nonumber \\
 \mbox{}  \hspace{10mm} - {\bf I}_b^\dagger \tilde{\bf d}_a(E_a,\varepsilon)
  - {\bf I}_a^\dagger \tilde{\bf d}_b(E_b,\varepsilon) = 0 .
\label{11}
\end{eqnarray}
Vector $ {\bf c}(0) $ occurring in Eq.~(\ref{8}) describes the
initial condition. If the electrons at both atoms are initially in
their ground states, $ {\bf c}^T(0) = [1,0,0,0] $.

Equations (\ref{8}---\ref{11}) can be solved as follows. First,
using Eq.~(\ref{11}) we express function $
\tilde{d}(E_a,E_b,\varepsilon) $ in the following form:
\begin{equation}   
 \tilde{d}(E_a,E_b,\varepsilon) = \frac{ {\bf I}_b^\dagger \tilde{\bf d}_a(E_a,\varepsilon)
  + {\bf I}_a^\dagger \tilde{\bf d}_b(E_b,\varepsilon) }{\varepsilon - E_a - E_b +
  2E_L} .
\label{12}
\end{equation}
Substitution of the expression (\ref{12}) into Eqs.~(\ref{9}) and
(\ref{10}) and subsequent integration in these equations results
in the following equations:
\begin{eqnarray}  
 \left[ (\varepsilon - E_a + E_L) {\bf 1} + {\bf L}_b \right]
  \tilde{\bf d}_a(E_a,\varepsilon) \nonumber \\
 \mbox{} \hspace{10mm} - \int dE_b \frac{ {\bf I}_b {\bf I}_a^\dagger
  \tilde{\bf d}_b(E_b,\varepsilon) }{ \varepsilon - E_a -E_b+2E_L}
  =  {\bf B}_a^\dagger \tilde{\bf c}(\varepsilon) ,
 \label{13} \\
 \left[ ( \varepsilon - E_b + E_L ) {\bf 1} + {\bf L}_a \right]
  \tilde{\bf d}_b(E_b,\varepsilon)  \nonumber \\
 \mbox{} \hspace{10mm} - \int dE_a \frac{{\bf I}_a {\bf I}_b^\dagger \tilde{\bf d}_a(E_a,\varepsilon)}{
  \varepsilon - E_a -E_b+2E_L}
  =  {\bf B}_b^\dagger \tilde{\bf c}(\varepsilon) .
\label{14}
\end{eqnarray}
The newly defined $ 2\times 2 $ matrices $ {\bf L}_a $ and $ {\bf
L}_b $ are given as
\begin{equation}  
 {\bf L}_j = - {\bf K}_j + i\pi {\bf I}_j {\bf I}_j^\dagger,
  \hspace{5mm} j=a,b.
\label{15}
\end{equation}

In the next step, we neglect the terms in Eqs.~(\ref{13}) and
(\ref{14}) containing the integrations over $ E_a $ and $ E_b $.
As shown in Appendix~A these terms equal zero for independent
atoms $ a $ and $ b $. Weakness of the dipole-dipole interaction
compared to the Coulomb one then justifies this approximation that
leaves us the following decoupled equations:
\begin{eqnarray}  
 \left[ (\varepsilon - E_a + E_L) {\bf 1} + {\bf L}_b \right]
  \tilde{\bf d}_a(E_a,\varepsilon)
  &=&  {\bf B}_a^\dagger \tilde{\bf c}(\varepsilon),
 \label{16} \\
 \left[ (\varepsilon - E_b + E_L) {\bf 1} + {\bf L}_a \right]
  \tilde{\bf d}_b(E_b,\varepsilon)
  &=&  {\bf B}_b^\dagger \tilde{\bf c}(\varepsilon) .
\label{17}
\end{eqnarray}
These equations can be solved using the decomposition of matrices
$ {\bf L}_a $ and $ {\bf L}_b $ \cite{PerinaJr2011},
\begin{equation} 
 {\bf L}_j = \sum_{k=1,2} \lambda_{k}^j {\bf L}_k^j,
  \hspace{5mm} j=a,b,
\label{18}
\end{equation}
in which $ \lambda^j_k $ are eigenvalues of the matrix $ {\bf L}_j
$ and matrices $ {\bf L}^j_1 $ and $ {\bf L}^j_2 $ fulfil the
relation $ {\bf L}^j_1 + {\bf L}^j_2 = {\bf 1} $. Formulas for
eigenvalues $ \lambda^j_k $ are derived as follows:
\begin{eqnarray} 
 & & \lambda^j_{1,2} = a^j_1 \pm \sqrt{ a^{j2}_1 - a^j_2 }, \hspace{10mm} j=a,b;
\label{19} \\
 & & \hspace{10mm} a^j_1 = \left[ i\pi \left(|\tilde\mu_j\alpha_L|^2
  + |V_j|^2 \right) - \Delta E_j^0 \right] / 2 , \nonumber \\
 & & \hspace{10mm} a^j_2 = - i\pi |\tilde\mu_j\alpha_L|^2
  \Delta E_j^0 - |\mu_j\alpha_L|^2 \nonumber  \\
 & & \hspace{20mm}  \mbox{} + i\pi \left(
  \tilde\mu_j^*\mu_j |\alpha_L|^2 V_j + {\rm c.c.}\right) ,
  \nonumber
\end{eqnarray}
where $ {\rm c.c.} $ denotes the complex conjugated term. Using
eigenvalues $ \lambda^j_k $, the matrices $ {\bf L}_k^j $ are
expressed as:
\begin{equation}  
 {\bf L}_1^j = \frac{1}{\lambda_1^j - \lambda_2^j} (
  {\bf L}_j - \lambda^j_2 {\bf 1} ), \hspace{5mm}
 {\bf L}_2^j = {\bf 1} -  {\bf L}_1^j .
\label{20}
\end{equation}
Substitution of the decomposition (\ref{18}) into Eq.~(\ref{16})
allows to obtain the solution of Eq.~(\ref{16}) in the simple
form:
\begin{equation}  
 \tilde{\bf d}_a(E_a,\varepsilon) = \sum_{k=1,2} \frac{ {\bf L}^b_k
  {\bf B}_a^\dagger \tilde{\bf c}(\varepsilon)}{
  \varepsilon - E_a + E_L + \lambda^b_k} .
\label{21}
\end{equation}
The solution of Eq.~(\ref{17}) is derived from that written in
Eq.~(\ref{21}) by the formal substitution $ a \leftrightarrow b $.

Substitution of the solutions $ \tilde{\bf d}_a(E_a,\varepsilon) $
and $ \tilde{\bf d}_b(E_b,\varepsilon) $ into Eq.~(\ref{8}) and
subsequent integration over energies $ E_a $ and $ E_b $ provides
the equation
\begin{equation} 
 \left[\varepsilon {\bf 1} - \bar{\bf A} \right] \tilde{\bf c}(\varepsilon)
  = i{\bf c}(0) .
\label{22}
\end{equation}
In Eq.~(\ref{22}), matrix $ \bar{\bf A} $ is defined as
\begin{equation}  
 \bar{\bf A} = {\bf A} -i\pi \sum_{j=a,b} {\bf B}_j
  {\bf B}_j^\dagger .
\label{23}
\end{equation}

The solution of Eq.~(\ref{22}) can be conveniently written in
terms of eigenvalues $ \Lambda_k $ and eigenvectors $ {\bf p}_k $
and $ {\bf p}^{\bf -1}_k $ of the matrix $ \bar{\bf A} $
decomposed as $ \bar{\bf A} = {\bf P} {\bf \Lambda} {\bf P^{-1}}
$,
\begin{equation} 
 \tilde{\bf c}(\varepsilon) = i \sum_{k=1}^{4} {\bf p}_k
  \frac{[{\bf p}^{\bf -1}_k {\bf c}(0)]}{ \varepsilon -\Lambda_k }  .
\label{24}
\end{equation}
Whereas eigenvectors $ {\bf p}_k $ form columns of the matrix $
{\bf P} $, eigenvectors $ {\bf p}^{\bf -1}_k $ occur as rows of
the inverse matrix $ {\bf P^{-1}} $.

Substituting Eq.~(\ref{24}) into Eq.~(\ref{21}), coefficients $
\tilde{\bf d}_a(E_a,\varepsilon) $ are obtained in the form:
\begin{equation} 
 \tilde{\bf d}_a(E_a,\varepsilon) = i \sum_{j=1}^{2} \sum_{k=1}^{4}
 \frac{ [{\bf L}^b_j{\bf B}^\dagger_a {\bf p}_k]}{ \varepsilon - E_a + E_L
  + \lambda_j^b} \, \frac{ [{\bf p}^{\bf -1}_k {\bf c}(0)] }{
  \varepsilon - \Lambda_k } .
\label{25}
\end{equation}
It can be shown by direct calculation that the integration over $
E_a $ in Eq.~(\ref{14}) using the solution (\ref{25}) gives zero.
Similarly, the solution for $ \tilde{\bf d}_b(E_b,\varepsilon) $
obtained from the symmetry $ a \leftrightarrow b $ results in zero
after the integration in Eq.~(\ref{13}). This confirms consistency
of the approximation leading to Eqs.~(\ref{16}) and (\ref{17}).

Finally, using Eq.~(\ref{25}) the expression (\ref{12}) for
coefficients $ \tilde{d}(E_a,E_b,\varepsilon) $ is expressed as:
\begin{eqnarray}   
 \tilde{d}(E_a,E_b,\varepsilon) = \frac{i}{\varepsilon - E_a - E_b + 2E_L}
   \Biggl\{ \sum_{j=1}^{2} \sum_{k=1}^{4} \nonumber \\
 \hspace{10mm} \mbox{}
  \frac{ [{\bf I}_b^\dagger {\bf L}^b_j {\bf B}_a^\dagger {\bf
  p}_k] }{ \varepsilon - E_a + E_L + \lambda_j^b }
  \frac{ [{\bf p}^{\bf -1}_k {\bf c}(0)] }{
  \varepsilon -\Lambda_k } + \{a\leftrightarrow b \} \Biggr\}.
\label{26}
\end{eqnarray}
Symbol $ \{a\leftrightarrow b \} $ in Eq.~(\ref{26}) replaces the
term obtained by the exchange of the indicated indices.

The inverse Laplace transform of the above formulas then provides
the coefficients in the time domain. Equation (\ref{24}) attains
the following form:
\begin{equation} 
 {\bf c}(t) = \sum_{k=1}^{4} {\bf p}_k [{\bf p}^{\bf -1}_k {\bf c}(0)]
  \exp(-i\Lambda_k t) .
\label{27}
\end{equation}
Decomposition of rational functions into partial fractions
followed by the Laplace transform leaves formula (\ref{25}) as
follows:
\begin{eqnarray} 
 & & {\bf d}_a(E_a,t) = \sum_{j=1}^{2} \sum_{k=1}^{4}
  \frac{[{\bf L}^b_j{\bf B}^\dagger_a {\bf p}_k] \, [{\bf p}^{\bf -1}_k{\bf c}(0)]
   }{ E_a - E_L - \lambda_j^b - \Lambda_k} \hspace{20mm} \nonumber \\
 & & \hspace{7mm} \mbox{} \times \left[ \exp[-i(E_a - E_L - \lambda_j^b)t]
  -  \exp(-i\Lambda_k t)  \right] .
\label{28}
\end{eqnarray}
Similarly, equation (\ref{26}) takes the following form in the
time domain:
\begin{eqnarray} 
 d(E_a,E_b,t) = \sum_{j=1}^{2} \sum_{k=1}^{4} \Biggl\{
  \frac{ [{\bf I}_b^\dagger {\bf L}^{b}_j{\bf B}_a^\dagger {\bf p}_k] \,
   [{\bf p}^{\bf -1}_k {\bf c}(0)] }{ E_a+E_b - 2E_L - \Lambda_k } \nonumber \\
 \mbox{} \hspace{10mm} \times \Biggl[ \frac{\exp(-i\Lambda_k t)-
  \exp[-i(E_a - E_L - \lambda_j^b)t] }{
  E_a - E_L - \lambda_j^b - \Lambda_k }  \nonumber \\
 \mbox{} \hspace{13mm} + \frac{ \exp[-i(E_a + E_b - 2E_L)t]}{
   E_b - E_L + \lambda_j^b } \nonumber \\
 \mbox{} \hspace{13mm}  - \frac{\exp[-i(E_a - E_L - \lambda_j^b)t] }{
  E_b - E_L + \lambda_j^b } \Biggr] + \{ a\leftrightarrow b \} \Biggr\} .
\label{29}
\end{eqnarray}

As eigenvalues $ \lambda_j^a $ and $ \lambda_j^b $ ($ \Lambda_k $)
have positive (negative) imaginary parts, the above formulas
considerably simplify in the long-time limit $ t \rightarrow
\infty $,
\begin{eqnarray} 
 {\bf c}^{\infty }(t) = [0,0,0,0]^T , \nonumber \\
 {\bf d}_j^{\infty }(E_j,t) = [0,0]^T, \hspace{10mm} j=a,b, \nonumber\\
 d^\infty(E_a,E_b,t) = \sum_{j=1}^{2} \sum_{k=1}^{4}
  \frac{ [{\bf p}^{\bf -1}_k {\bf c}(0)] }{ E_a+E_b - 2E_L - \Lambda_k } \nonumber \\
 \mbox{} \hspace{10mm} \times \Biggl\{ \frac{ [{\bf I}_b^\dagger
  {\bf L}^{b}_j{\bf B}_a^\dagger {\bf p}_k] }{ E_b - E_L + \lambda_j^b }
   + \frac{ [{\bf I}_a^\dagger
  {\bf L}^{a}_j{\bf B}_b^\dagger {\bf p}_k] }{ E_a - E_L + \lambda_j^a }
  \Biggr\} \nonumber \\
 \mbox{} \hspace{10mm} \times \exp[-i(E_a + E_b - 2E_L)t] .
\label{30}
\end{eqnarray}
Both atoms are thus completely ionized in the long-time limit. In
this limit, also the norm of the long-time joint photoelectron
ionization spectrum $ |d^\infty(E_a,E_b)|^2 $ is determined
analytically,
\begin{eqnarray}   
 \int dE_a \int dE_b |d^\infty(E_a,E_b)|^2 = 4\pi^2
  \sum_{j,j'=1}^{2} \sum_{k,k'=1}^{4}
  \nonumber \\
 \hspace{10mm} \Biggl\{ \frac{ [{\bf I}_b^\dagger{\bf L}^{b}_j{\bf B}_a^\dagger {\bf
  p}_k]^* [{\bf I}_b^\dagger {\bf L}^{b}_{j'}{\bf B}_a^\dagger {\bf
  p}_{k'}] }{ \lambda^{b*}_j - \lambda^b_{j'} } +
  \{ a\leftrightarrow b \} \Biggr\} \nonumber \\
 \hspace{10mm} \mbox{} \times  \frac{ [{\bf p}^{\bf -1}_k {\bf c}(0)]^*
  [{\bf p}^{\bf -1}_{k'} {\bf c}(0)] }{ \Lambda^*_{k} -
  \Lambda_{k'} }.
\label{31}
\end{eqnarray}

As both atoms interact through the dipole-dipole interaction,
energies $ E_a $ and $ E_b $ of the ionized electrons are
correlated to some extent. This correlation is quantified by
covariance $ C $ defined as:
\begin{equation}   
 C = \frac{ \langle E_a E_b \rangle }{ \sqrt{ \langle
 E_a^2\rangle \langle E_b^2\rangle } }.
\label{32}
\end{equation}
In Eq.~(\ref{32}), the mean values $ \langle E_a^k E_b^l \rangle $
for $ k,l \in N $ are defined as:
\begin{equation}  
 \langle E_a^k E_b^l \rangle = \int dE_a \int dE_b \,
  E_a^k E_b^l |d^\infty(E_a,E_b)|^2
\label{33}
\end{equation}
using the long-time joint photoelectron ionization spectrum $
|d^\infty(E_a,E_b)|^2 $ determined in Eq.~(\ref{30}). As these
correlations have quantum origin, they can alternatively be
described by quadratic negativity as it is done in the following
section.

\section{Quadratic negativity for the composite system}

We first consider a simplified model analyzed in
\cite{PerinaJr2011,PerinaJr2011a}. In this model, atom $ a $ has
only two discrete states $ |0\rangle_a $ and $ |1\rangle_a $
whereas atom $ b $ contains the whole continuum $ |E_b\rangle $ of
states with energies $ E_b $. A general state $ |\psi_q\rangle $
of this system described by coefficients $ d_0(E_b) $ and $
d_1(E_b) $ takes the following form:
\begin{equation}  
 |\psi_q\rangle = \int dE_b \, \left[d_0(E_b)|0\rangle_a +
  d_1(E_b)|1\rangle_a \right] |E_b\rangle .
\label{34}
\end{equation}
Quadratic negativity $ N_{q} $ quantifies the amount of
non-separability of a state \cite{Hill1997} as visible in
negativity of the corresponding partially transposed statistical
operator \cite{Peres1996,Horodecki1996}. It is determined by
negative eigenvalues of the partially transposed statistical
operator. Quadratic negativity of the state $ |\psi_q\rangle $
written in Eq.~(\ref{34}) has been derived in \cite{Luks2012} in
the form:
\begin{eqnarray}   
 N_q &=& 2 \sum_{j,k=0}^{1} \int dE_b \int dE'_b
  \left[  d_j^*(E_b) d_j(E_b) d_k^*(E'_b) d_k(E'_b)
  \right. \nonumber \\
 & & \left. \mbox{}
   - d_j^*(E_b) d_k(E_b) d_k^*(E'_b) d_j(E'_b) \right] .
\label{35}
\end{eqnarray}

Formula~(\ref{35}) can be recast as follows:
\begin{equation}   
 N_q = 2 \int dE_b \varrho_b(E_b) \int dE'_b \varrho_b(E'_b)
   \, n^q(E_b,E'_b) ,
\label{36}
\end{equation}
where $ \varrho_b $ gives the density of states $ |E_b) $ in the
continuum of atom $ b $:
\begin{equation}    
 \varrho_b(E_b) = \sum_{j=0}^{1} |d_j(E_b)|^2 .
\label{37}
\end{equation}
Joint density of quadratic negativity $ n_q $ introduced in
Eq.~(\ref{36}) is obtained in the form:
\begin{eqnarray}  
 n_q(E_b,E'_b) = 1 - \frac{ \sum_{j,k=0}^{1} d^*_j(E_b)d_k(E_b)
  d^*_k(E'_b)d_j(E'_b)}{\varrho_b(E_b)\varrho_b(E'_b)}. \nonumber
  \\
\label{38}
\end{eqnarray}
The joint density of quadratic negativity $ n_q $ gives quadratic
negativity of the bipartite system formed by the states $ \{
|0\rangle_a, |1\rangle_a \} $ and $ \{ |E_b) , |E'_b) \} $.
According to Eq.~(\ref{36}), quadratic negativity $ N_q $ is given
as a weighted sum of quadratic negativities between the two-level
atom $ a $ and all possible pairs of states inside the continuum
of atom $ b $.

This interpretation suggests a straightforward generalization to
the case of two continua substituting $ |0\rangle_a \rightarrow
|E_a) $ and $ |1\rangle_a \rightarrow |E'_a) $ and integrating
over frequencies $ E_a $ and $ E'_a $:
\begin{eqnarray}   
 N = \int dE_a \int dE'_a \int dE_b \int dE'_b \nonumber \\
 \mbox{} \hspace{10mm}  \left[  d^*(E_a,E_b) d(E_a,E_b)
  d^*(E'_a,E'_b) d(E'_a,E'_b) \right. \nonumber \\
  \mbox{} \hspace{10mm}  \left.
   - d^*(E_a,E_b) d(E'_a,E_b) d^*(E'_a,E'_b) d(E_a,E'_b) \right] .
\label{39}
\end{eqnarray}
Coefficients $ d(E_a,E_b) $ determine a state $ |\psi\rangle $
common to both continua at atoms $ a $ and $ b $:
\begin{equation}  
 |\psi\rangle = \int dE_a \int dE_b \, d(E_a,E_b) |E_a) |E_b) .
\label{40}
\end{equation}
They also give densities $ \varrho_a $ and $ \varrho $ describing
an ionized electron at atom $ a $ and both ionized electrons,
respectively:
\begin{eqnarray} 
 \varrho(E_a,E_b) = |d(E_a,E_b)|^2 ,\nonumber \\
 \varrho_a(E_a) = \int dE_b \, \varrho(E_a,E_b) .
\label{41}
\end{eqnarray}

Following the same argumentation as that applied earlier for the
simplified model considered in \cite{Luks2012}, the joint density
of quadratic negativity $ n $ can be introduced:
\begin{eqnarray}   
 n(E_a,E'_a,E_b,E'_b) = 2 \left[ \varrho_{ab}(E_a,E_b)
   \varrho_{ab}(E'_a,E'_b) \right. \nonumber \\
 \mbox{} \hspace{10mm} \left. \mbox{} - d^*(E_a,E_b) d(E'_a,E_b) d^*(E'_a,E'_b) d(E_a,E'_b)
   \right] \nonumber \\
 \mbox{} \hspace{10mm} \times \left\{ \varrho_a(E_a) \varrho_a(E'_a)
   [\varrho_{ab}(E_a,E_b)+\varrho_{ab}(E'_a,E_b)] \right. \nonumber \\
 \mbox{} \hspace{10mm} \left. \mbox{} \times [\varrho_{ab}(E_a,E'_b)+
   \varrho_{ab}(E'_a,E'_b)] \right\}^{-1} .
\label{42}
\end{eqnarray}
Quadratic negativity $ N $ can then be expressed using the
densities $ \varrho_a $ and $ \varrho $ introduced in
Eq.~(\ref{41}):
\begin{eqnarray}   
 N = \int dE_a \varrho_a(E_a) \int dE'_a \varrho_a(E'_a) \nonumber \\
 \mbox{} \hspace{10mm} \times \int dE_b [\varrho(E_a,E_b)+\varrho(E'_a,E_b)] \nonumber \\
 \mbox{} \hspace{10mm} \times \int dE'_b [\varrho(E_a,E'_b)+\varrho(E'_a,E'_b)]
  \nonumber \\
 \mbox{} \hspace{10mm} \times n(E_a,E'_a,E_b,E'_b) .
 \label{43}
\end{eqnarray}

Formula~(\ref{43}) can be recast into a simple form using the
reduced statistical operator $ \tilde\rho_b $ of the continuum of
atom $ b $:
\begin{eqnarray}  
 N = 2 \left[ 1 - \int dE_b \int dE'_b\, |\tilde\rho_b(E_b,E'_b)|^2 \right] , \\
 \tilde\rho_b(E_b,E'_b) = \int dE_a \, d(E_a,E_b) d^*(E_a,E'_b) .
\end{eqnarray}
An analogous formula based on the reduced statistical operator of
atom $ a $ can also be derived.

Another expression for quadratic negativity $ N $ can be reached
using the Schmidt decomposition of state $ |\psi\rangle $ given in
Eq.~(\ref{40}):
\begin{equation}  
  d(E_a,E_b) = \sum_n f_n(E_a) \lambda_n g_n(E_b) ;
\label{46}
\end{equation}
$ \lambda_n $ being coefficients of the decomposition. Functions $
f_n $ and $ g_n $ introduced in Eq.~(\ref{46}) form the dual
orthonormal bases. Quadratic negativity $ N $ can then be
determined according to the formula:
\begin{equation}  
 N = 2\left[ 1 -\sum_n \lambda_n^4 \right].
\label{47}
\end{equation}
Any separable state gives $ N = 0 $, whereas we have $ N =
2(N_d-1)/N_d $ for the maximally entangled state in $ N_d \times
N_d $ dimensions. We note that the formula in Eq.~(\ref{47}) can
be rewritten as $ N = 2 ( 1 - 1/K ) $ where $ K $ denotes the
Schmidt number of independent modes.

We note that if atoms $ a $ and $ b $ contain also discrete levels
the consideration of amplitude spectra with $ \delta $-functions
allows for easy incorporation of such levels into the above
developed description (for details, see \cite{Luks2012}).

Entanglement of two ionized electrons can easily be modified by
filtering the energies of electrons. Quadratic negativity $ N_a $
characterizing a state with energies $ E_a $ of electron $ a $ in
interval $ \langle E_a^0-\Delta E,E_a^0+\Delta E \rangle $ is
obtained along the formula:
\begin{eqnarray}   
 N_a(E_a^0) = {\cal N}_a^{-2}\int_{E_a^0-\Delta E}^{E_a^0+\Delta
 E} dE_a \varrho_a(E_a) \int_{E_a^0-\Delta E}^{E_a^0+\Delta
 E} dE'_a  \nonumber \\
 \mbox{} \hspace{10mm} \varrho_a(E'_a) \int dE_b [\varrho_{ab}(E_a,E_b)+\varrho_{ab}(E'_a,E_b)] \nonumber \\
 \mbox{} \hspace{10mm} \times \int dE'_b
 [\varrho_{ab}(E_a,E'_b)+\varrho_{ab}(E'_a,E'_b)]\nonumber \\
 \mbox{} \hspace{13mm} \times
  n(E_a,E'_a,E_b,E'_b) ,
\label{48}  \\
{\cal N}_a = \int_{E_a^0-\Delta E}^{E_a^0+\Delta
  E} dE_a  \int dE_b \, |d(E_a,E_b)|^2 . \nonumber
\end{eqnarray}

If also energies $ E_b $ of electron $ b $ are limited to interval
$ \langle E_b^0-\Delta E,E_b^0+\Delta E \rangle $, quadratic
negativity $ N_{ab} $ of the resultant state is determined as
follows:
\begin{eqnarray}   
 N_{ab}(E_a^0,E_b^0) = {\cal N}^{-2} \int_{E_a^0-\Delta E}^{E_a^0+\Delta
  E} dE_a \varrho_a(E_a) \int_{E_a^0-\Delta E}^{E_a^0+\Delta
  E} dE'_a \nonumber \\
 \mbox{} \hspace{10mm}  \varrho_a(E'_a) \int_{E_b^0-\Delta E}^{E_b^0+\Delta E} dE_b
  [\varrho_{ab}(E_a,E_b)+\varrho_{ab}(E'_a,E_b)] \nonumber \\
 \mbox{} \hspace{10mm} \times \int_{E_b^0-\Delta E}^{E_b^0+\Delta E} dE'_b
  [\varrho_{ab}(E_a,E'_b)+\varrho_{ab}(E'_a,E'_b)] \nonumber \\
 \mbox{} \hspace{13mm} \times  n(E_a,E'_a,E_b,E'_b) ,
\label{49}  \\
{\cal N} = \int_{E_a^0-\Delta E}^{E_a^0+\Delta
  E} dE_a  \int_{E_b^0-\Delta E}^{E_b^0+\Delta E} dE_b \, |d(E_a,E_b)|^2 . \nonumber
\end{eqnarray}

\section{Long-time photoelectron ionization spectra}

In the following, we restrict ourselves to the most common case of
two identical atoms. The discussion of photoelectron ionization
spectra is divided into two parts according to the relative
strength of direct and indirect ionization paths. For the
discussion, we introduce a useful parametrization that generalizes
the parametrization introduced by Fano
\cite{Fano1961,PerinaJr2011b}. In our parametrization, parameters
$ \mu_j $, $ \tilde\mu_j $, $ V_j $, and $ J_j $ ($j=a,b$) are
replaced by the following parameters \cite{PerinaJr2011a}:
\begin{eqnarray}   
 \gamma_j = \pi |V_j|^2, \hspace{5mm}
  \bar\gamma_j = \pi |J_j|^2, & \nonumber \\
 q_j = \frac{\mu_j}{\pi\tilde\mu_j V_j^*}, \hspace{5mm}
  \bar{q}_a = \frac{\mu_b}{\pi\tilde\mu_a J_a^*}, \hspace{5mm}
  \bar{q}_b = \frac{\mu_a}{\pi\tilde\mu_b J_b^*}, & \nonumber \\
 \Gamma_j = \gamma_j + \bar\gamma_j, \hspace{5mm} Q_j = \frac{\gamma_j q_j + \bar\gamma_j \bar q_j}{\Gamma_j}, &
  \nonumber \\
 \Omega_j = \sqrt{4\pi\Gamma_j}(Q_j+i) \tilde\mu_j \alpha_L, &
  \hspace{-15mm} j=a,b.
\label{50}
\end{eqnarray}
In Eq.~(\ref{50}), $ \gamma_j $ ($ \bar\gamma_j $) gives damping
of the continuum at atom $ j $ caused by the Coulomb
(dipole-dipole) interaction and $ \Gamma_j $ is the overall
damping. Parameter $ q_j $ ($ \bar{q}_j $) gives the ratio of
indirect and direct ionization strengths related to the Coulomb
(dipole-dipole) interaction. Parameter $ \Omega_j $ is linearly
proportional to the pumping strength of ionization at atom $ j $.
As both atoms interact with the same laser field of amplitude $
\alpha_L $, the introduction of two additional parameters is
convenient:
\begin{eqnarray}   
 & m = \frac{\mu_b}{\mu_a} , \hspace{5mm}
 \Omega = \frac{\Omega_a+\Omega_b}{2}.
\label{51}
\end{eqnarray}
Whereas parameter $ m $ gives the ratio of dipole moments for the
auto-ionizing states at both atoms, parameter $ \Omega $ describes
average pumping strength. A suitable parametrization is based upon
common parameters $ m $ and $ \Omega $ and parameters $ \gamma_j
$, $ \bar\gamma_j $ and $ q_j $ of individual atoms defined in
Eq.~(\ref{50}). We note that $ m=1 $ for two identical atoms and
parameters $ \gamma_j $, $ q_j $ and $ \Omega_j = \Omega $, $
j=a,b $, coincide with the usual ones defined by Fano
\cite{Fano1961}.

\subsection{Spectra for comparable direct and indirect ionization paths}

\begin{figure}        
 \begin{center}
 (a) \resizebox{0.7\hsize}{!}{\includegraphics{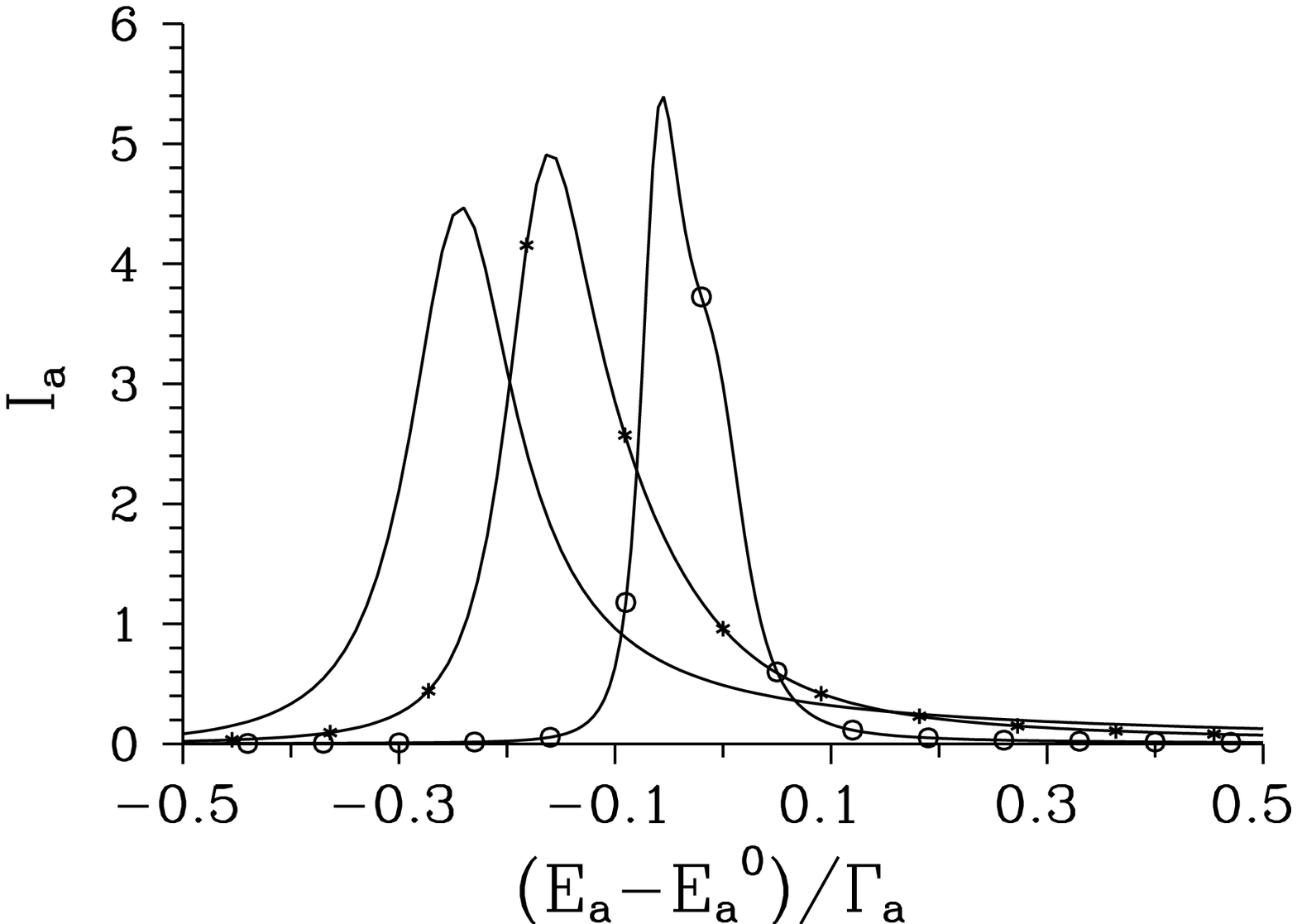}}

 \vspace{2mm}
 (b) \resizebox{0.7\hsize}{!}{\includegraphics{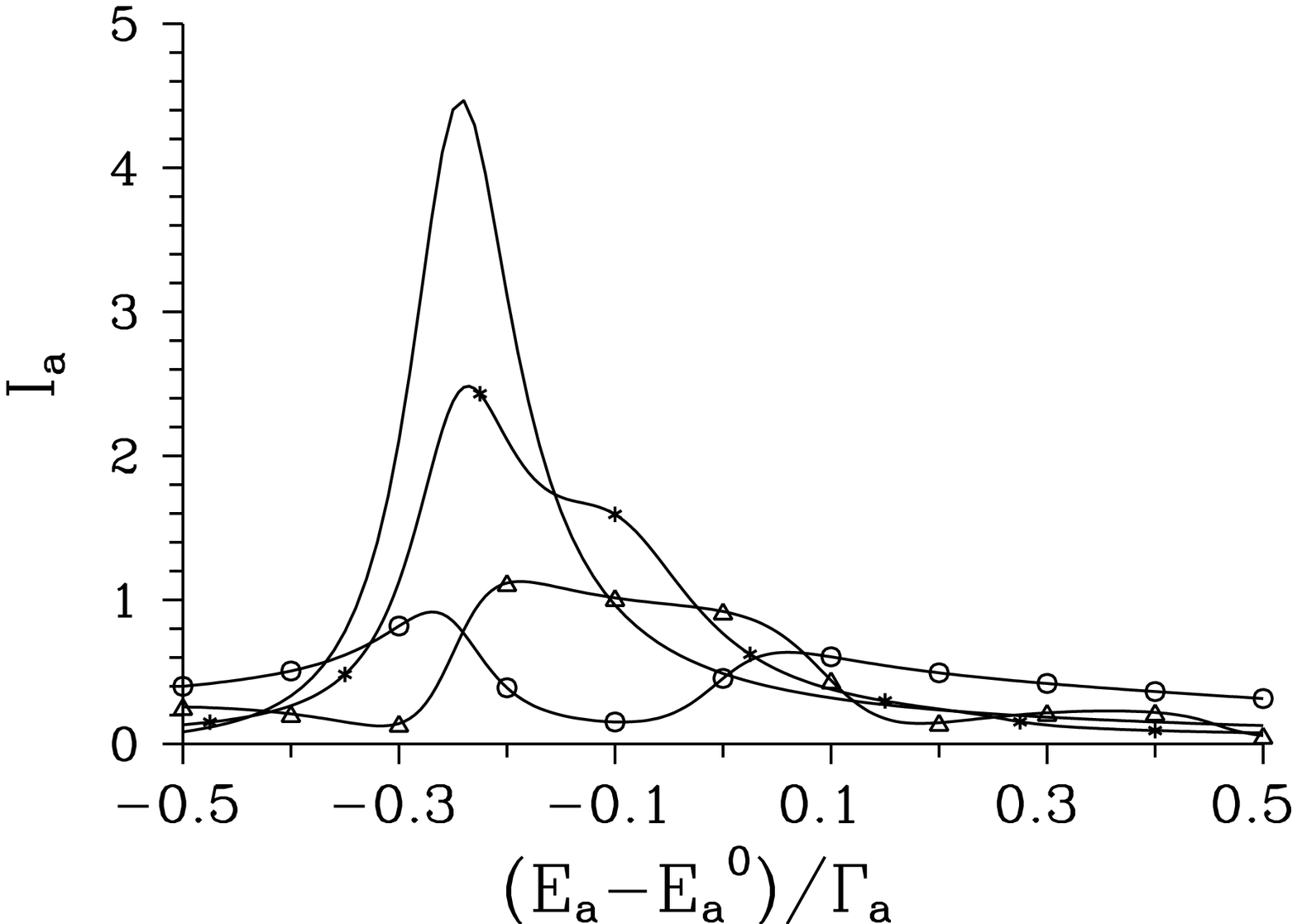}}
 \end{center}
 \caption{Long-time intensity photoelectron ionization spectrum $ I_a $
  of atom $ a $ [$ I_a(E_a) = \int dE_b \, |d^\infty(E_a,E_b)|^2 $]
  as a function of normalized energy $ (E_a-E_a^0)/\Gamma_a $ for
  (a) $ \bar\gamma_a = \bar\gamma_b  = 0 $ (independent atoms, curve without symbols),
  0.1 (curve with $ \ast $), 1 (curve with $ \circ $), $ J_{ab} = 0 $
  and (b) $ J_{ab} = 0 $ (independent atoms, curve without symbols),
  0.56 (curve with $ \ast $), 1 (curve with $ \circ $),
  1.68 (curve with $ \triangle $), $ \bar\gamma_a = \bar\gamma_b = 0 $;
  $ E_a = E_b = E_L = 1 $, $ q_a = q_b = 1 $, $ \gamma_a = \gamma_b = 1 $,
  $ \Omega = 1 $, $ m = 1 $.}
\label{fig2}
\end{figure}

If the values of the Fano parameters $ q_a $ and $ q_b $ are close
to the unity they characterize the case in which both ionization
paths compete each other. This results in the creation of the Fano
zeroes at energies $ E^0_a -\gamma_a q_a $ and $ E^0_b -\gamma_b
q_b $ of electrons at atoms $ a $ and $ b $, respectively.
Individual spectra are formed by peaks that move down to lower
energies with the increasing values of pumping parameter $ \Omega
$. Whereas the peaks are above the Fano energy for lower values of
pumping parameter $ \Omega $, they occur below the Fano energy for
greater values of parameter $ \Omega $ \cite{Fano1961}. As shown
in Fig.~\ref{fig2}(a) for ionization spectrum $ I_a $ of atom $ a
$, the dipole-dipole interaction with continuum (described by
parameters $ J_a $ and $ J_b $) moves the peaks towards energies $
E_a^0 $ and $ E_b^0 $ of the auto-ionizing levels. This behavior
reflects the fact that the dipole-dipole interaction with
continuum itself ionizes atoms $ a $ and $ b $ and thus diminishes
the contributions of direct and indirect ionization paths whose
interference is responsible for energy shifts of the peaks. On the
other hand, the dipole-dipole interaction between the discrete
auto-ionizing levels (described by parameter $ J_{ab} $) tends to
form a peak close to energy $ E_a^0 $ (or $ E_b^0 $) in individual
ionization spectra, as documented in Fig.~\ref{fig2}(b). Two peaks
coexist together in the spectra for intermediate values of
parameter $ J_{ab} $. If the dipole-dipole interaction between
discrete auto-ionizing levels is sufficiently strong, there occurs
one large peak in the photoelectron ionization spectrum $ I_a $ of
atom $ a $ (or atom $ b $). It is separated by two minima from the
tails built by the direct and indirect ionization paths. The
presence of two minima in the spectrum reflects nontrivial mutual
influence between the dipole-dipole interaction on one side and
the direct and indirect ionization paths on the other side.

\begin{figure}        
 \begin{center}
 (a) \resizebox{0.7\hsize}{!}{\includegraphics{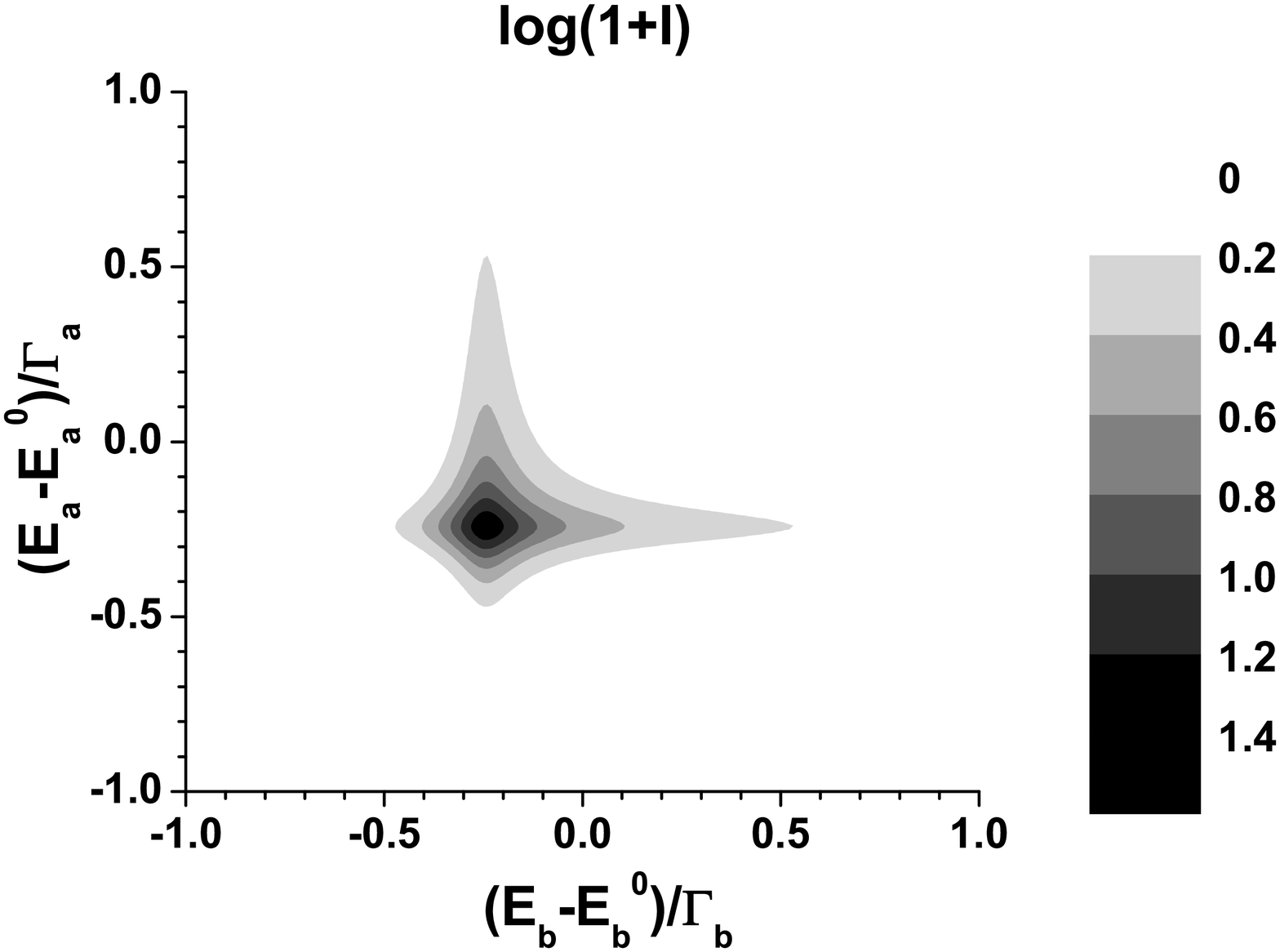}}

 \vspace{4mm}
 (b) \resizebox{0.7\hsize}{!}{\includegraphics{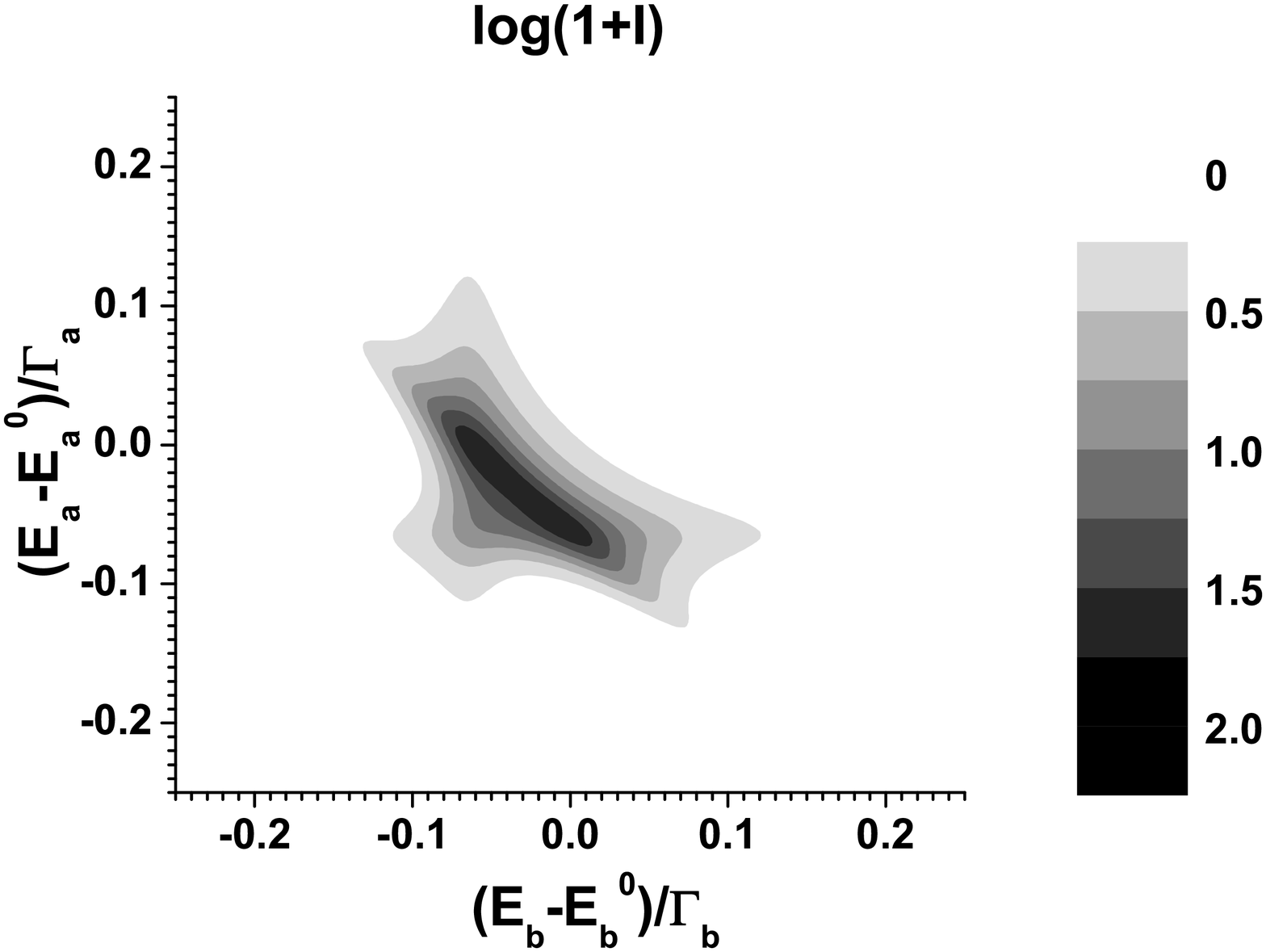}}

 \vspace{4mm}
 (c) \resizebox{0.7\hsize}{!}{\includegraphics{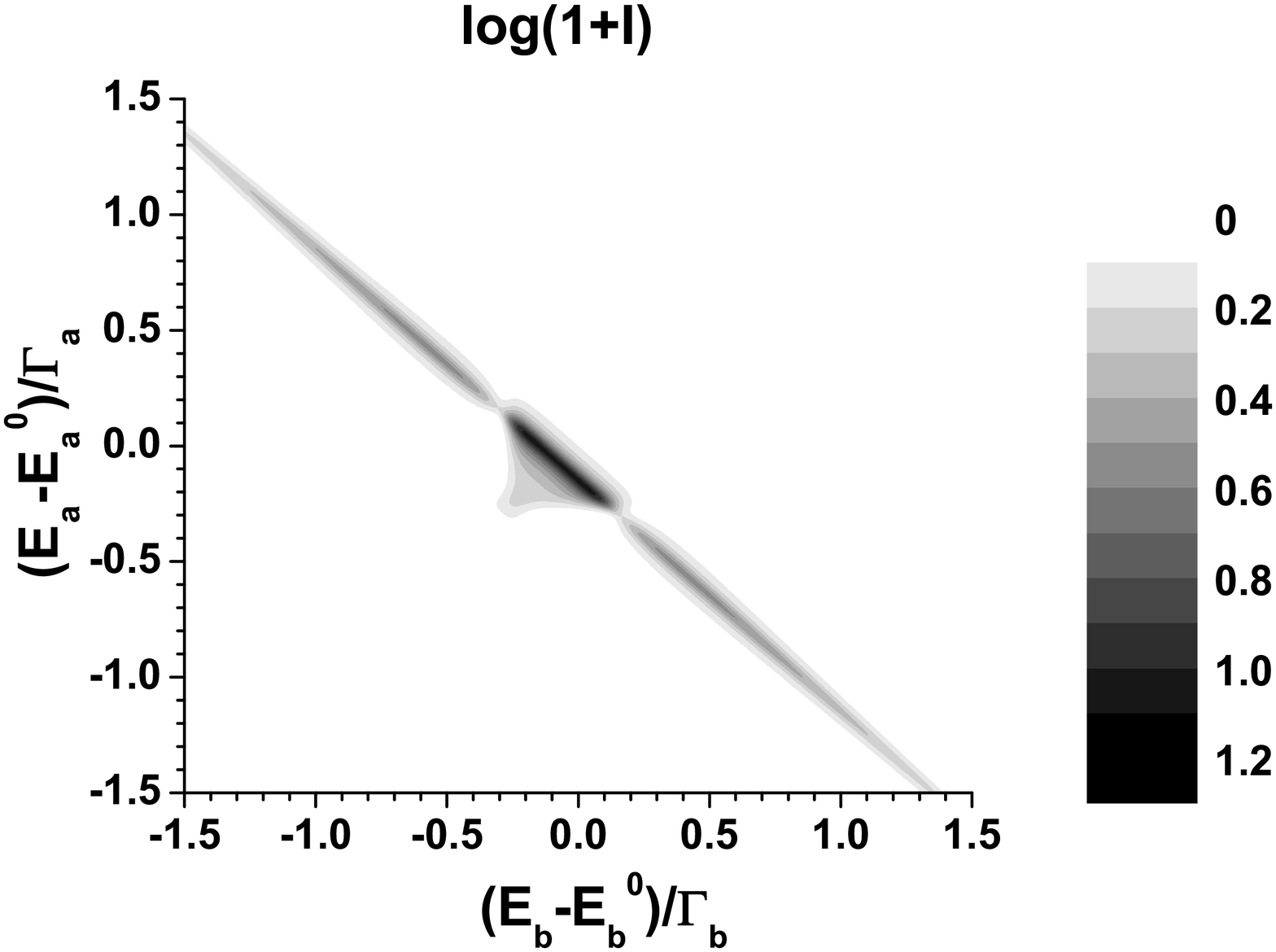}}
 \end{center}
 \caption{Topo graphs of long-time joint photoelectron
  ionization spectra $ I $ plotted as a function of normalized energies
  $ (E_a-E_a^0)/\Gamma_a $  and
  $ (E_b-E_b^0)/\Gamma_b $ of atoms $ a $ and $ b $, respectively, for
  (a) $ \bar\gamma_a = \bar\gamma_b  = 0 $, $ J_{ab} = 0 $ (independent atoms),
  (b) $ \bar\gamma_a = \bar\gamma_b  = 1 $, $ J_{ab} = 0 $ and
  (c) $ \bar\gamma_a = \bar\gamma_b  = 0 $, $ J_{ab} = 1.68 $.
  Values of the other parameters are written in the caption to
  Fig.~\ref{fig2}; $ \log $ stands for decimal logarithm.}
\label{fig3}
\end{figure}

The dipole-dipole interaction mutually correlates energies of the
ionized electrons at atoms $ a $ and $ b $ (see Fig.~\ref{fig3}).
This results in pure states of two ionized electrons entangled in
energies $ E_a $ and $ E_b $. Correlations increase with the
increasing strength of both dipole-dipole interactions with the
continua and between the discrete auto-ionizing levels. In fact,
there occurs 'anti-correlation' between energies $ E_a $ and $ E_b
$ of the electrons that expresses the conservation law of energy
in the stationary system ($ E_a + E_b \approx 2E_L $). Comparison
of the joint photoelectron ionization spectra $ I $ plotted in
Figs.~\ref{fig3}(b) and (c) reveals that the stronger the
dipole-dipole interaction the tighter the energy correlations.

\begin{figure}        
 \begin{center}
 (a) \resizebox{0.7\hsize}{!}{\includegraphics{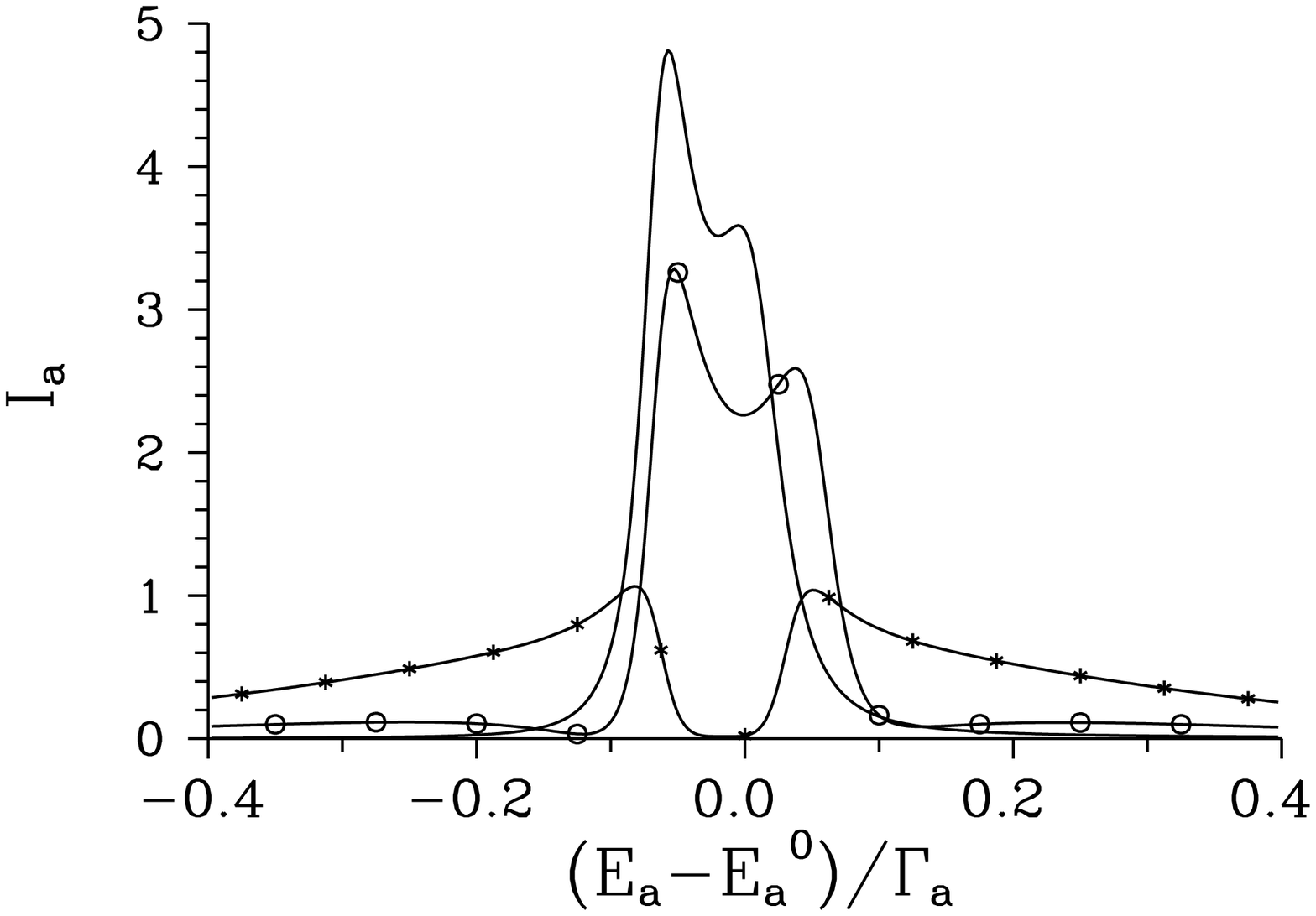}}

 \vspace{2mm}
 (b) \resizebox{0.7\hsize}{!}{\includegraphics{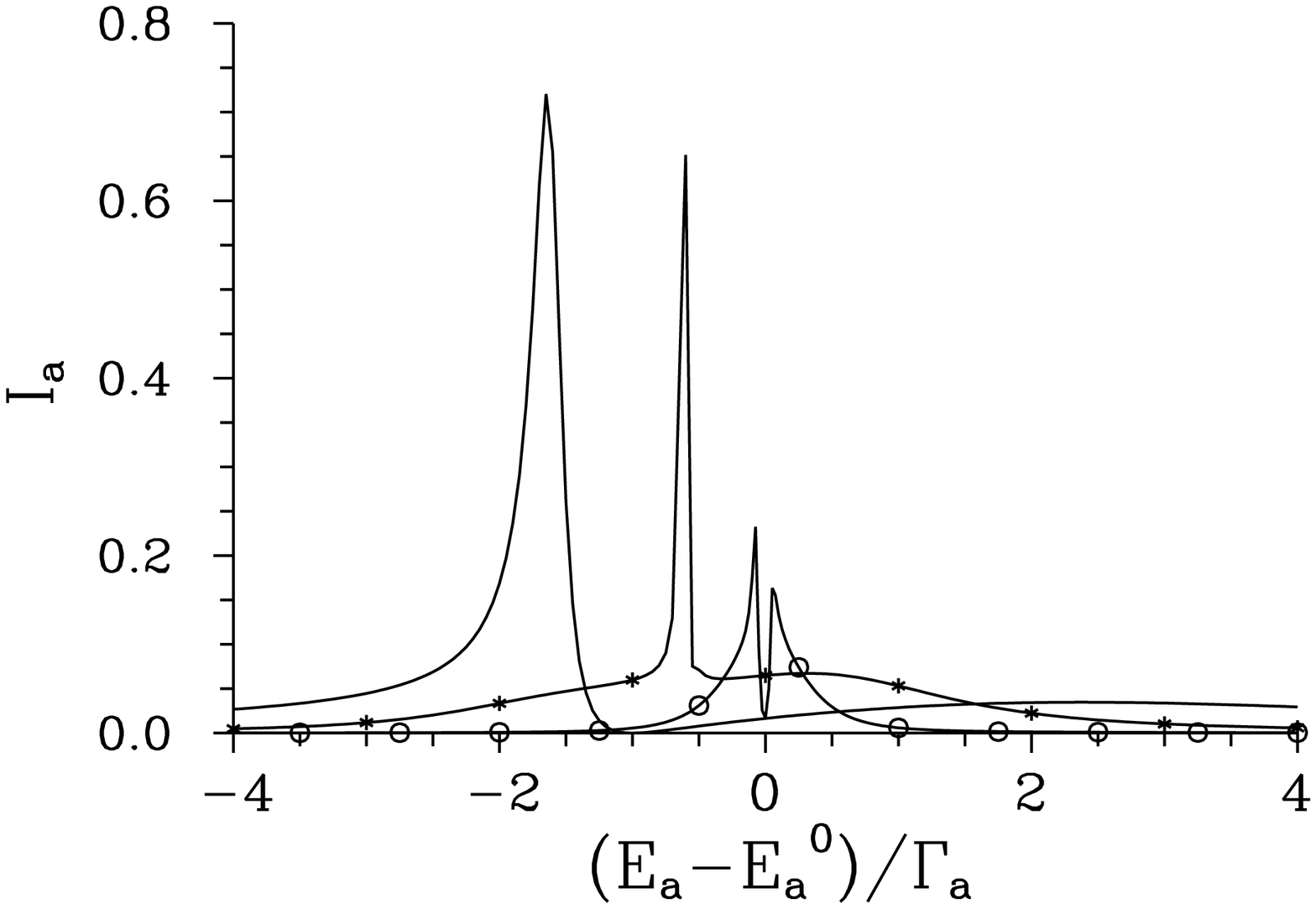}}
 \end{center}
 \caption{Long-time photoelectron ionization spectrum $ I_a $
  of atom $ a $ for (a) $ J_{ab} = 0.56 $ (curve without symbols),
  2 (curve with $ \ast $), 4 (curve with $ \circ $),
  $ \bar\gamma_a = \bar\gamma_b = 1 $, $ \Omega = 1 $
  and (b) $ \bar\gamma_a = \bar\gamma_b = 0 $, $ J_{ab} = 0 $
  (independent atoms, curve without symbols), $ \bar\gamma_a = \bar\gamma_b = 1 $,
  $ J_{ab} = 0 $ (curve with $ \ast $), $ \bar\gamma_a = \bar\gamma_b = 1 $,
  $ J_{ab} = 2 $ (curve with $ \circ $), $ \Omega = 5 $. Values of
  the other parameters are written in the caption to
  Fig.~\ref{fig2}. Curves with $ \bar\gamma_a = \bar\gamma_b = 1
  $ and $ J_{ab} = 2 $ fulfil the 'balance condition' given in Eq.~(\ref{B10}).}
\label{fig4}
\end{figure}

If the dipole-dipole interactions with the continuum and between
the discrete auto-ionizing levels have comparable strengths, a
well-formed minimum occurs in the photoelectron ionization
spectrum $ I_a $ ($ I_b $) of atom $ a $ ($ b $) (see
Fig.~\ref{fig4}). If the original spectra without the
dipole-dipole interactions lie above the Fano zeros, intensity of
this minimum is close to zero [see Fig.~\ref{fig4}(a)]. If the
original spectra are below the Fano zeros, the minimum intensity
does not reach zero but the intensity profile exhibits a sharp
minimum accompanied by a neighbor sharp peak [see
Fig.~\ref{fig4}(b)]. Both cases give a clear evidence about the
strong mutual influence of ionization channels based on the
continuum and the discrete levels. Conditions for the balance of
two ionization channels based on the dipole-dipole interaction
have been found in Appendix~B using the Fano diagonalization
approach.

\subsection{Spectra for the dominating indirect ionization path}

\begin{figure}        
 \centerline{\resizebox{0.7\hsize}{!}{\includegraphics{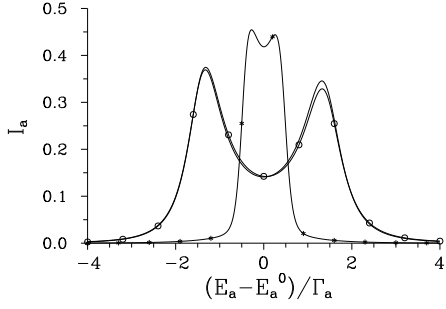}}}

 \caption{Long-time photoelectron ionization spectrum $ I_a $
  of atom $ a $ for $ \bar\gamma_a = \bar\gamma_b = 0 $,
  $ J_{ab} = 0 $ (independent atoms, curve without symbols),
  $ \bar\gamma_a = \bar\gamma_b = 1 $,
  $ J_{ab} = 0 $ (curve with $ \ast $) and
  $ \bar\gamma_a = \bar\gamma_b = 0 $,
  $ J_{ab} = 0.56 $ (curve with $ \circ $);
  $ E_a = E_b = E_L = 1 $, $ q_a = q_b = 100 $, $ \gamma_a = \gamma_b = 1 $,
  $ \Omega = 3 $, $ m = 1 $.}
\label{fig5}
\end{figure}

High values of the Fano parameters $ q_a $ and $ q_b $ occur in
this region, in which the ionized states are reached nearly
exclusively from discrete auto-ionizing levels with energies $
E_a^0 $ and $ E_b^0 $. The individual photoelectron ionization
spectra of both atoms thus consist of peaks centered around energy
$ E_a^0 $ ($ E_b^0$) for atom $ a $ ($ b $). There occurs the
Autler-Townes splitting of these peaks \cite{Autler1955} for
intense pumping. It occurs whenever the speed of populating an
auto-ionizing level is faster than its depletion due to the
Coulomb configuration interaction ($ |\mu_a \alpha_L| > |V_a| $).
As shown in Fig.~\ref{fig5} comparing the cases of independent and
interacting atoms, the dipole-dipole interaction with the
continuum weakens the Autler-Townes splitting. This is a
consequence of the fact that the dipole-dipole interaction makes
the transfer of electrons from discrete excited auto-ionizing
levels into their continua faster and so partly suppresses the
effect of 'population reversion' at the discrete auto-ionizing
levels responsible for the splitting. On the other hand, the
dipole-dipole interaction between the discrete auto-ionizing
levels only slightly modifies the ionization spectra, as
documented in Fig.~\ref{fig5}. This follows from the considered
symmetric configuration of atoms $ a $ and $ b $ that minimizes
the influence of parameter $ J_{ab} $ to the populations of
discrete levels $ E_a^0 $ and $ E_b^0 $.

\begin{figure}        
 \begin{center}
 (a) \resizebox{0.7\hsize}{!}{\includegraphics{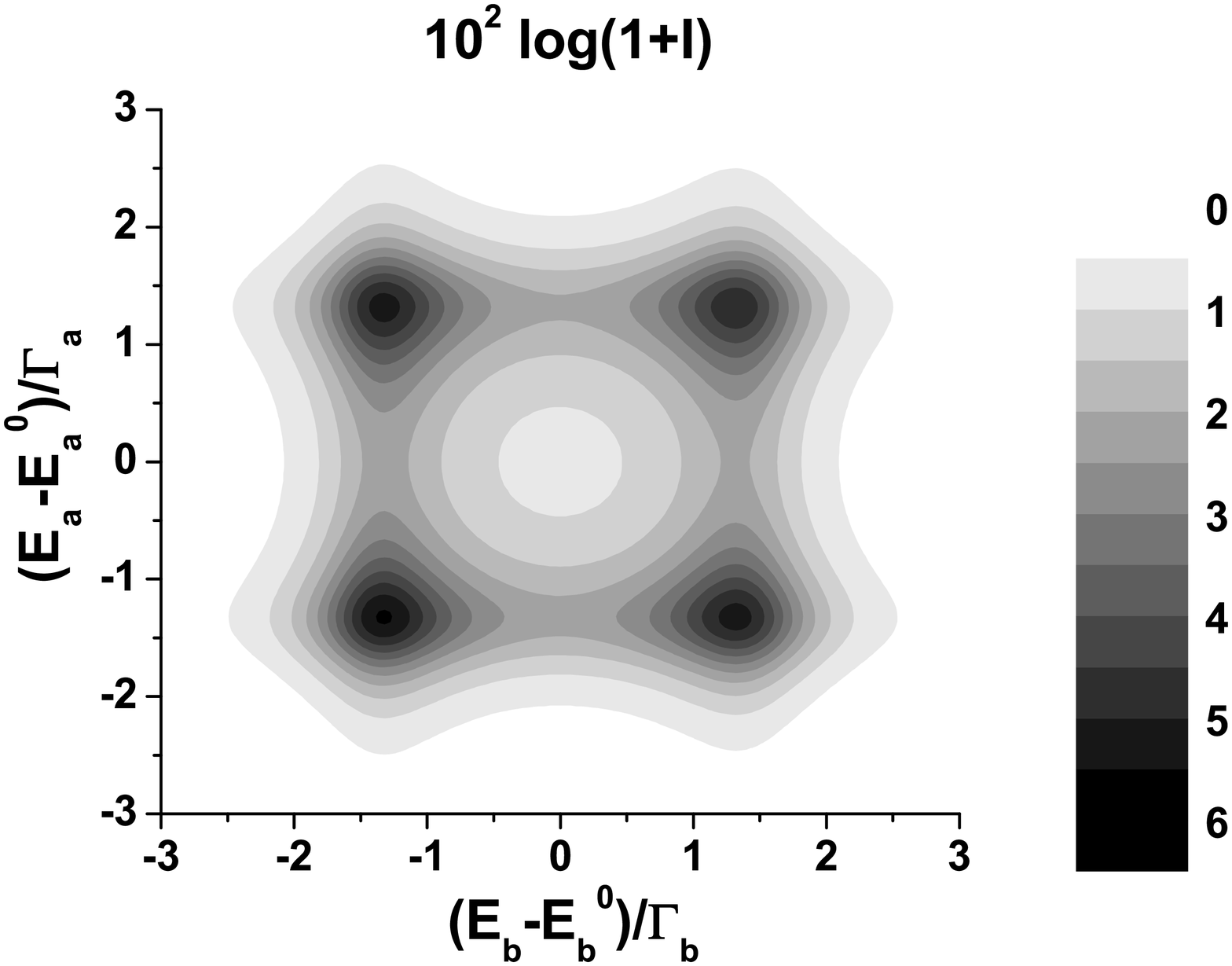}}

 \vspace{4mm}
 (b) \resizebox{0.7\hsize}{!}{\includegraphics{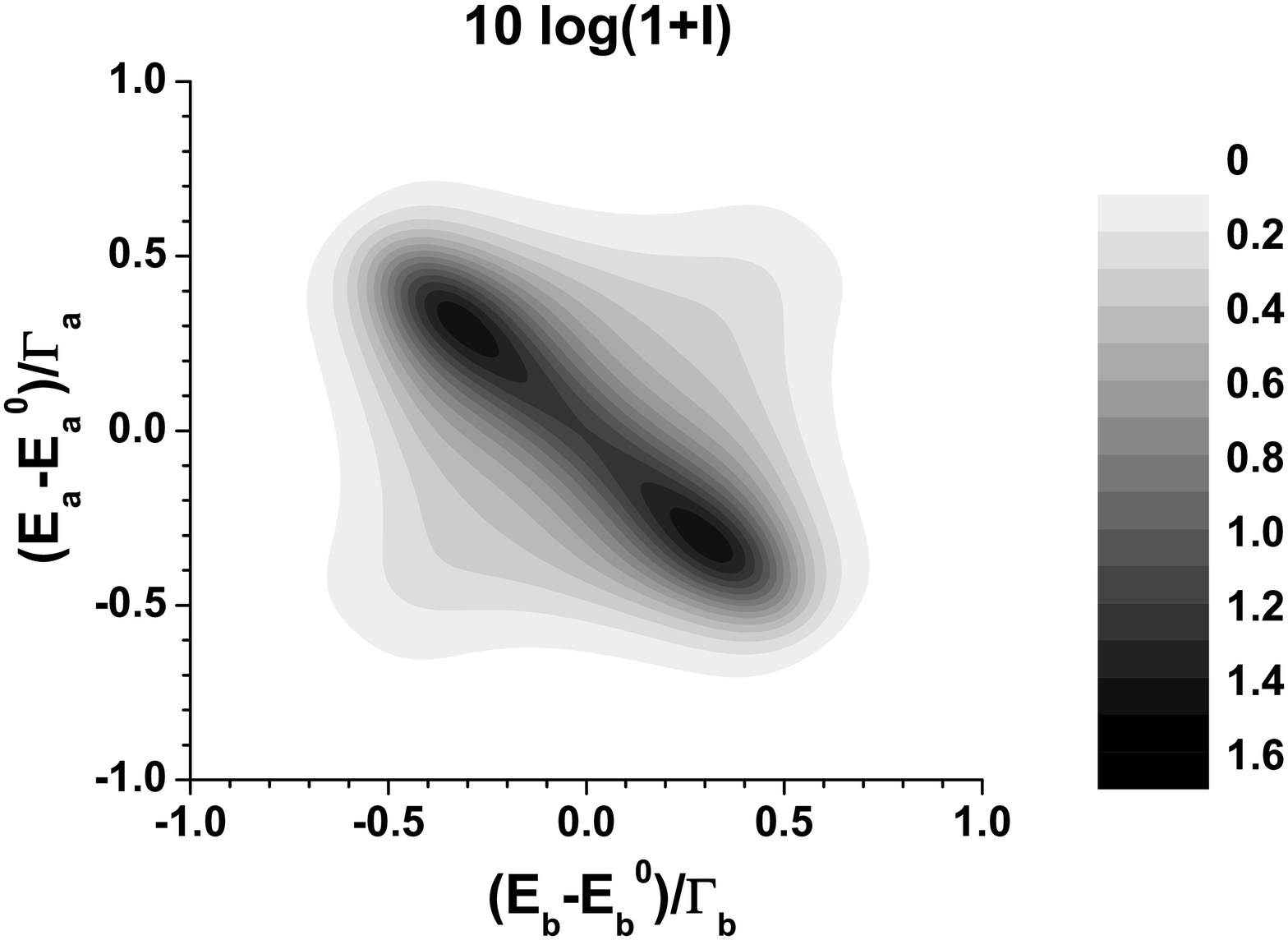}}
 \end{center}
 \caption{Topo graphs of long-time joint photoelectron
  ionization spectra $ I $ for
  (a) $ \bar\gamma_a = \bar\gamma_b  = 0 $ (independent atoms) and
  (b) $ \bar\gamma_a = \bar\gamma_b  = 1 $; $ J_{ab} = 0 $ and
  values of the other parameters are written in the caption to
  Fig.~\ref{fig5}.}
\label{fig6}
\end{figure}

The dipole-dipole interaction with the continuum strongly modifies
the joint photoelectron ionization spectra $ I $ that belong to
the states entangled in energies. Dramatic change of the joint
photoelectron ionization spectrum $ I $ caused by this interaction
is illustrated in Fig.~\ref{fig6} in which the dipole-dipole
interaction nearly completely suppresses the Autler-Townes
splitting.

\section{Entanglement in long-time photoelectron ionization
spectra}

The joint photoelectron ionization spectra $ I $ plotted in
Figs.~\ref{fig3}(b) and (c) and \ref{fig6}(b) reflect strong
correlations in energies $ E_a $ and $ E_b $ of the ionized
electrons created due to the dipole-dipole interaction. These
correlations emerging during the quantum evolution are
non-classical. They reflect the bipartite entanglement that can be
quantified by quadratic negativity $ N $ introduced in Sec.~3.
Alternatively, they can be described by the von Neumann entropy $
S $ of the reduced statistical operators of individual electrons $
a $ and $ b $. We note that entropy $ S $ is a monotonous function
of negativity $ N $. Contrary to entropy, quadratic negativity $ N
$ can be expressed via its joint density $ n $ introduced in
Eq.~(\ref{41}). This represents an important advantage as it
allows to connect entanglement with spectral properties of the
ionized electrons.

\begin{figure}        
 \begin{center}
 \resizebox{0.45\hsize}{!}{\includegraphics{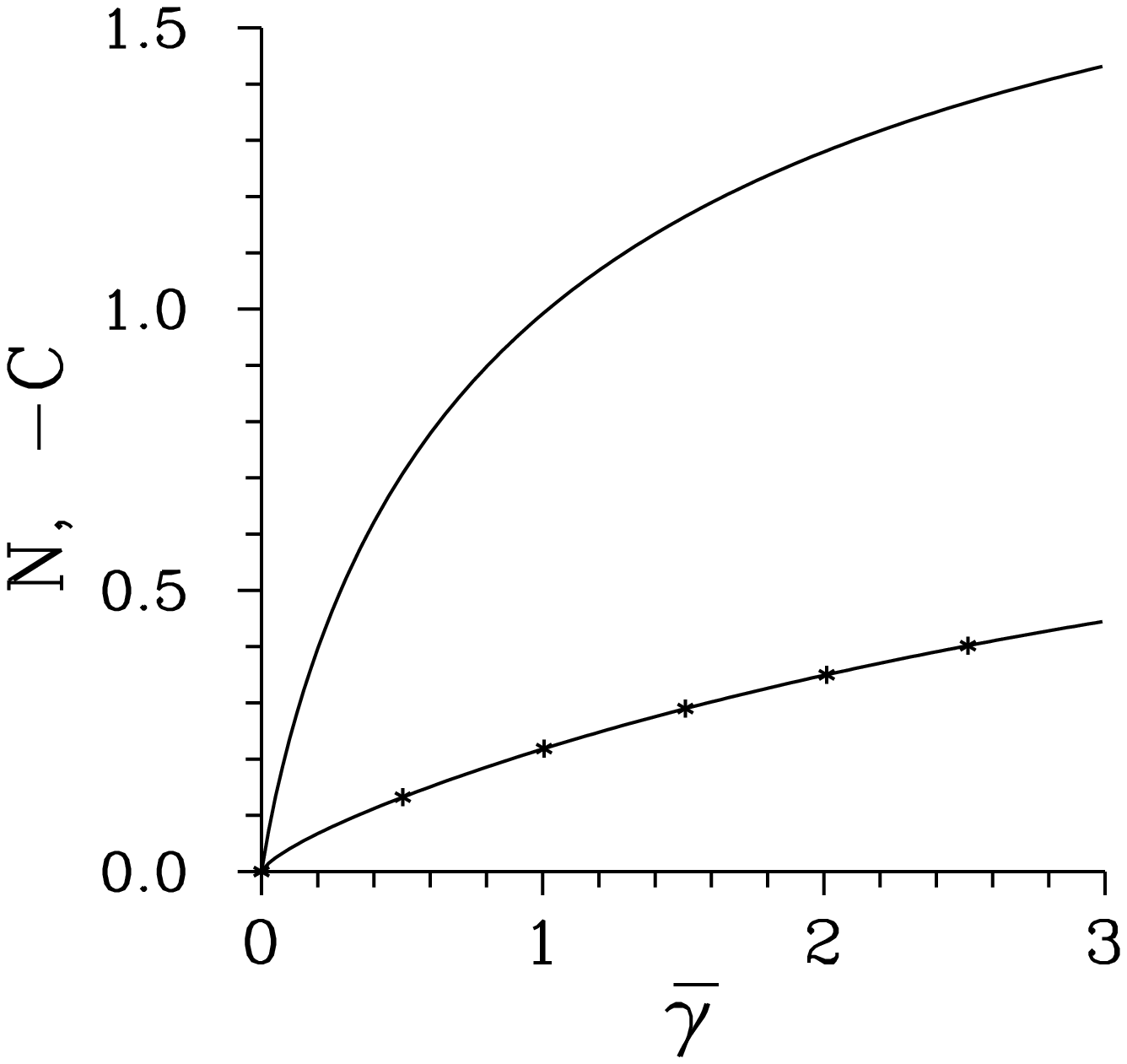}}
 \hspace{1mm}
 \resizebox{0.45\hsize}{!}{\includegraphics{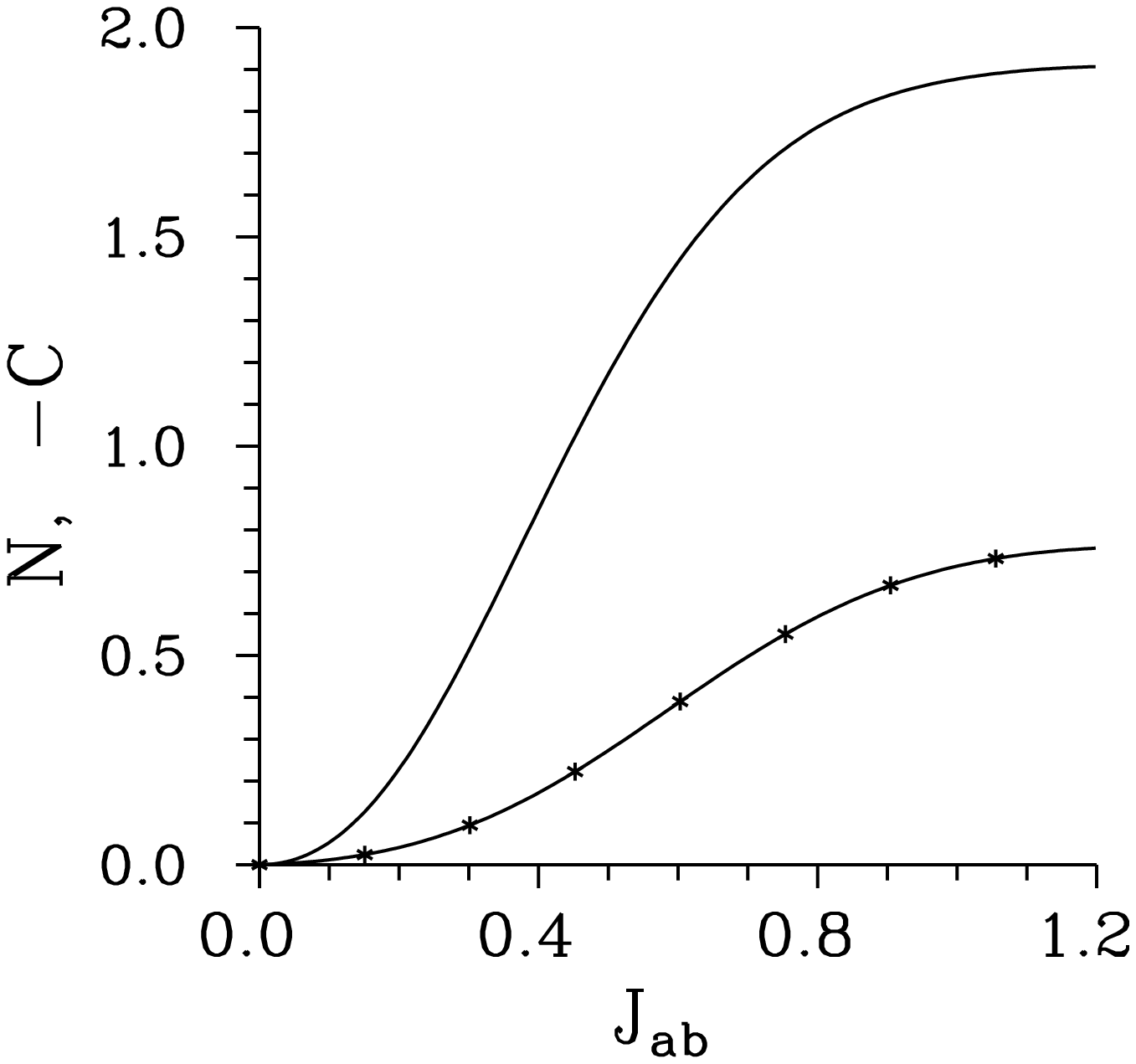}}

 \centerline{(a) \hspace{.4\hsize} (b)}
 \end{center}
 \caption{Quadratic negativity $ N $ (curve without symbols) and
  covariance $ C $ (curve with $ \ast $) as they depend on (a) parameter
  $ \bar\gamma \equiv \bar\gamma_a = \bar\gamma_b $ and (b)
  parameter $ J_{ab} $. Values of the other parameters are written
  in the caption to Fig.~\ref{fig2}.}
\label{fig7}
\end{figure}

Assuming other parameters fixed, the increasing values of
parameters $ J_a $ ($ \tilde\gamma_a $) and $ J_b $ ($
\bar\gamma_b $) describing the dipole-dipole interactions with the
continua result in greater values of negativity $ N $ [see
Fig.~\ref{fig7}(a)]. Similarly, the greater the values of
parameter $ J_{ab} $ characterizing the dipole-dipole interaction
between the discrete auto-ionizing levels, the greater the values
of negativity $ N $ [see Fig.~\ref{fig7}(b)]. In Fig.~\ref{fig7},
the comparison of curves giving negativity $ N $ and covariance $
C $ of energies reveals that even the classical covariance $ C $
can be used as a good indicator of mutual coupling of two ionized
electrons.

\begin{figure}        
 \centerline{\resizebox{0.7\hsize}{!}{\includegraphics{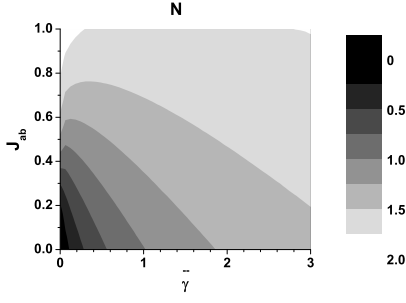}}}

 \caption{Topo graph of quadratic negativity $ N $ as it depends
  on dipole-dipole interaction parameters $ \bar\gamma \equiv \bar\gamma_a =
  \bar\gamma_b $ and $ J_{ab} $. Values of the parameters are written
  in the caption to Fig.~\ref{fig2}.}
\label{fig8}
\end{figure}

Whereas the dipole-dipole interactions with the continua and
between the discrete auto-ionizing levels influence the ionization
spectra in general in a complex way, they support each other in
creating entanglement, as documented in Fig.~\ref{fig8}.

\begin{figure}        
 \begin{center}
 (a) \resizebox{0.7\hsize}{!}{\includegraphics{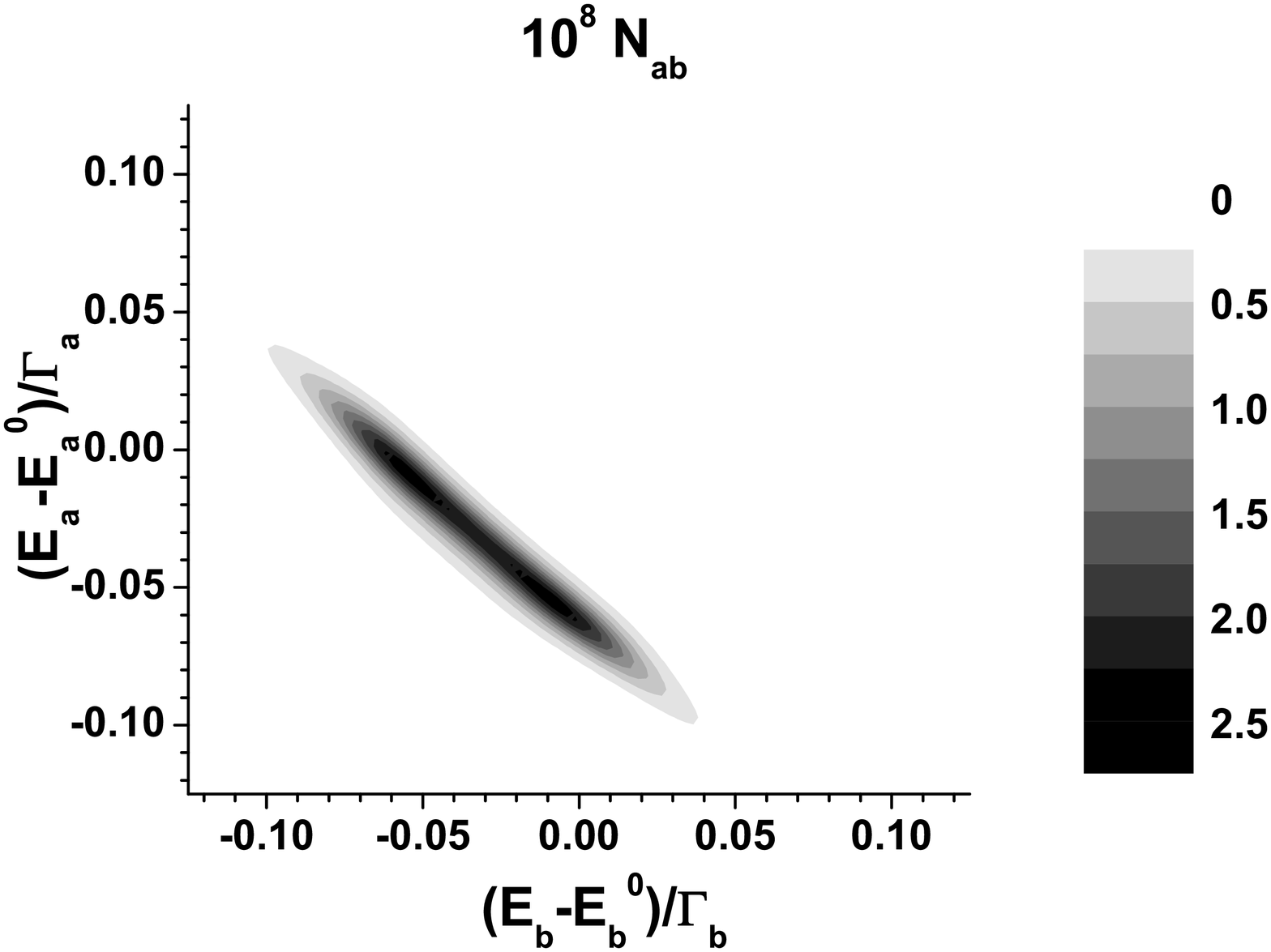}}

 \vspace{4mm}
 (b) \resizebox{0.7\hsize}{!}{\includegraphics{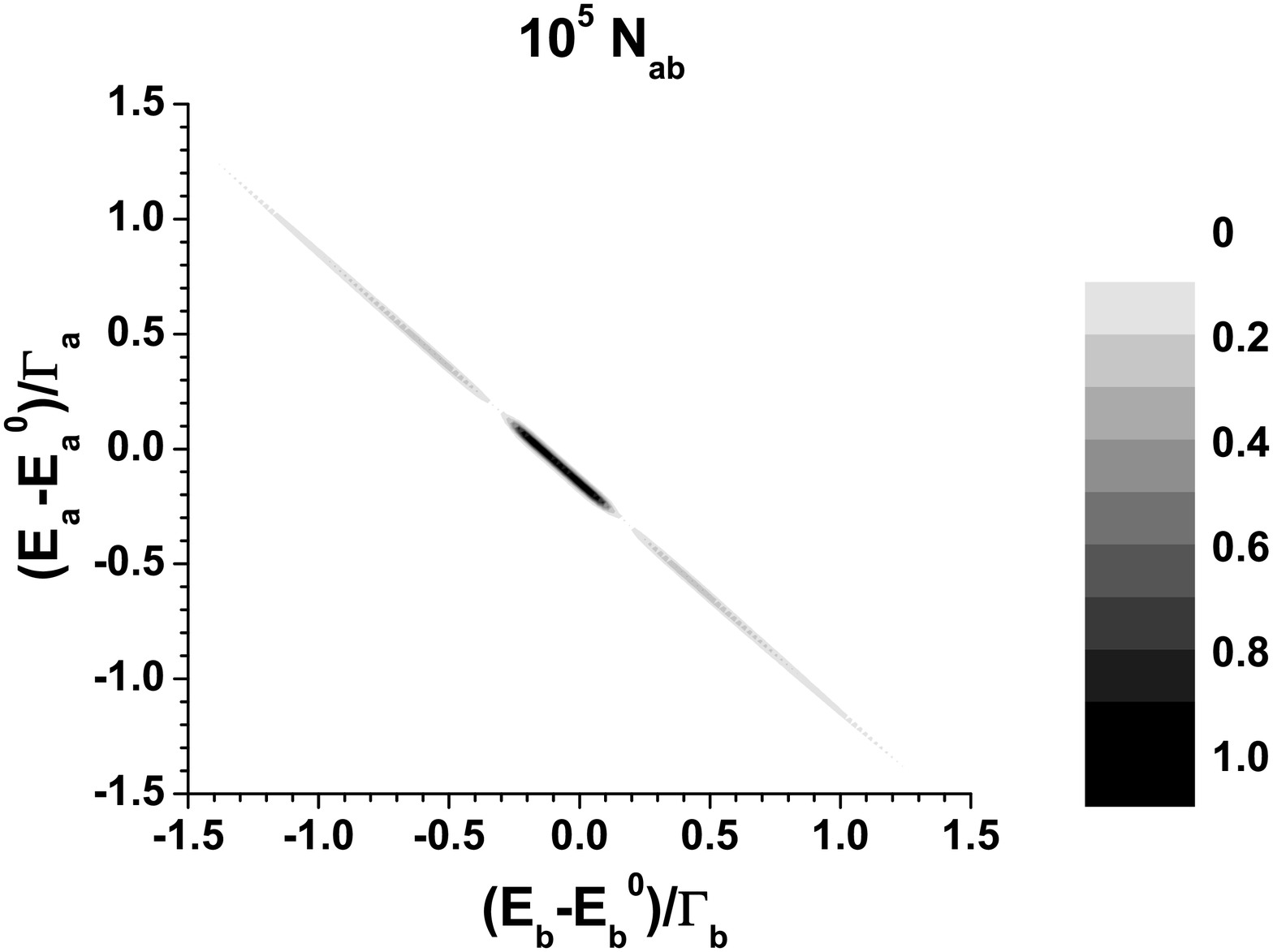}}
 \end{center}
 \caption{Topo graphs of quadratic negativity $ N_{ab} $ depending
  on normalized energies $ (E_a-E_a^0)/\Gamma_a $  and
  $ (E_b-E_b^0)/\Gamma_b $ for
  (a) $ \bar\gamma_a = \bar\gamma_b  = 1 $, $ J_{ab} = 0 $, $ \Delta E = 0.005 $
  [$ N = 0.98 $] and (b) $ \bar\gamma_a = \bar\gamma_b  = 0 $, $ J_{ab} = 1.68 $,
  $ \Delta E = 0.01 $ [$ N = 1.79 $].
  Values of the other parameters are written in the caption to
  Fig.~\ref{fig2}.}
\label{fig9}
\end{figure}

Spectral distribution of entanglement can be conveniently
visualized using negativity $ N_{ab} $ defined in Eq.~(\ref{49}).
It characterizes a common state of both ionized electrons obtained
by filtering the energies of electrons. The comparison of graphs
shown in Figs.~\ref{fig9}(a) and \ref{fig6}(b) and also in
Figs.~\ref{fig9}(b) and \ref{fig3}(c) reveals that the negativity
$ N $ is concentrated in the areas with higher intensities.

\begin{figure}        
 \centerline{\resizebox{0.7\hsize}{!}{\includegraphics{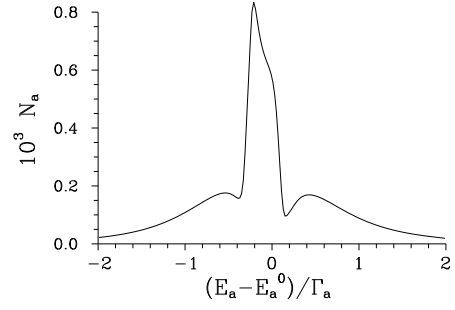}}}

 \caption{Quadratic negativity $ N_a $ as a function of
  normalized energy $ (E_a-E_a^0)/\Gamma_a $ of atom $ a $;
  $ \Delta E = 0.01 $, $ J_{ab} = 1.68 $ and values of the other parameters are
  written in the caption to Fig.~\ref{fig2}; $ N = 1.79 $.}
\label{fig10}
\end{figure}

Also when the energy of only one electron is filtered, highly
entangled states are obtained in the central part of the spectrum
(see Fig.~\ref{fig10} for negativity $ N_a $). We note that the
values of negativity $ N_{ab} $ as well as negativity $ N_a $
depend on the length $ \Delta E $ of the interval of measured
energies [see Eqs.~(\ref{48}) and (\ref{49})]. The wider the
interval $ \Delta E $ the greater the values of negativities.
However, this dependence is weak.

\section{Ionization in molecular condensates}

The dipole-dipole interaction is usually much weaker than the
dipole interaction of atoms or molecules with external coherent
fields. It is also much weaker than the Coulomb configuration
interaction. This means that only weak modifications of the
ionization spectra discussed in Sec.~4 are expected in real
systems. For example, energy shifts of the ionization peaks in the
spectra plotted in Fig.~\ref{fig2} are comparable to the value of
the dipole-dipole interaction energy. On the other hand, strongly
correlated pairs of ionized electrons can be obtained in real
systems provided that the process of ionization is sufficiently
slow compared to the timescale characterizing the dipole-dipole
interaction.

\begin{figure}        
 \begin{center}
 \resizebox{0.45\hsize}{!}{\includegraphics{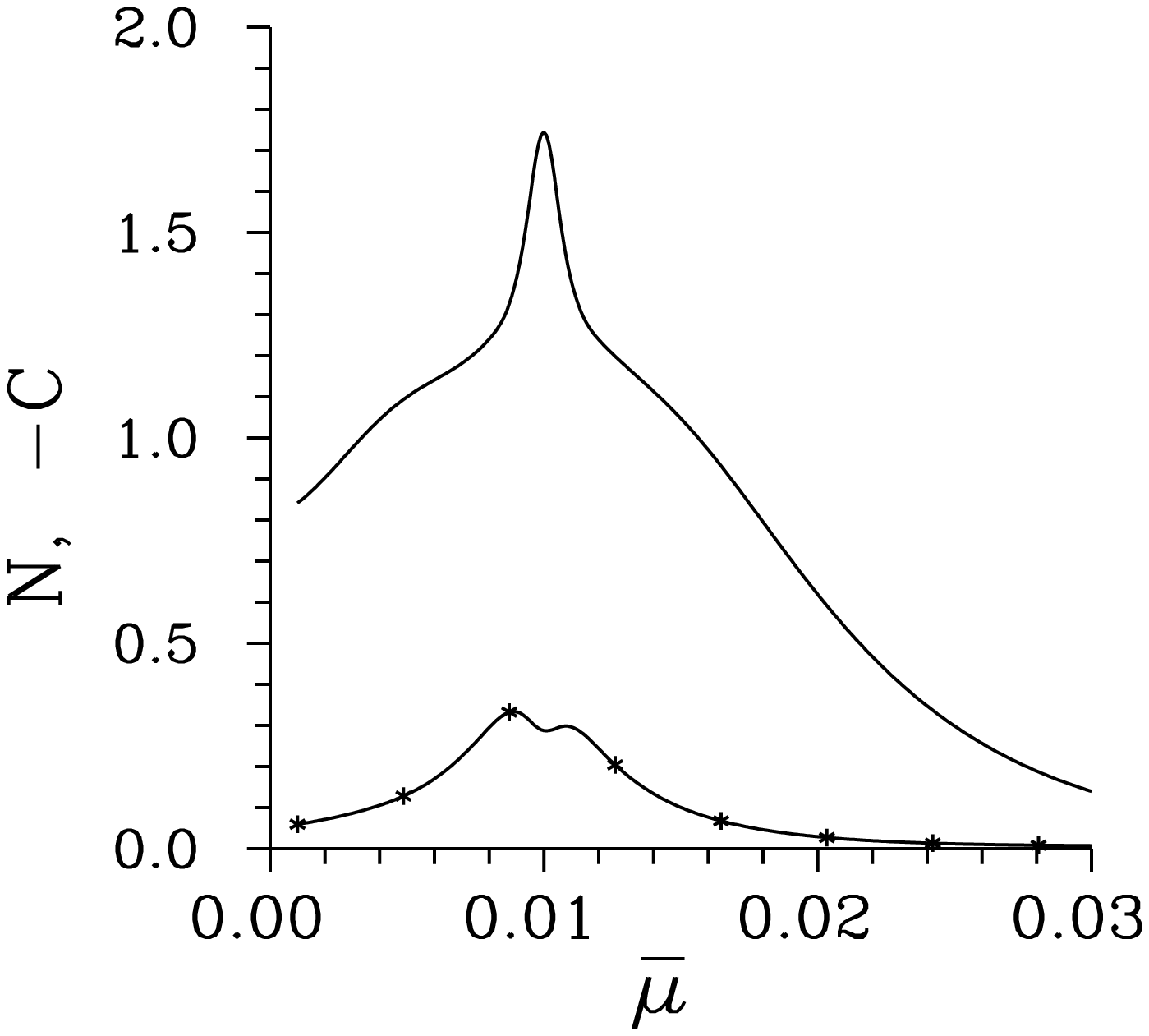}}
 \hspace{1mm}
 \resizebox{0.45\hsize}{!}{\includegraphics{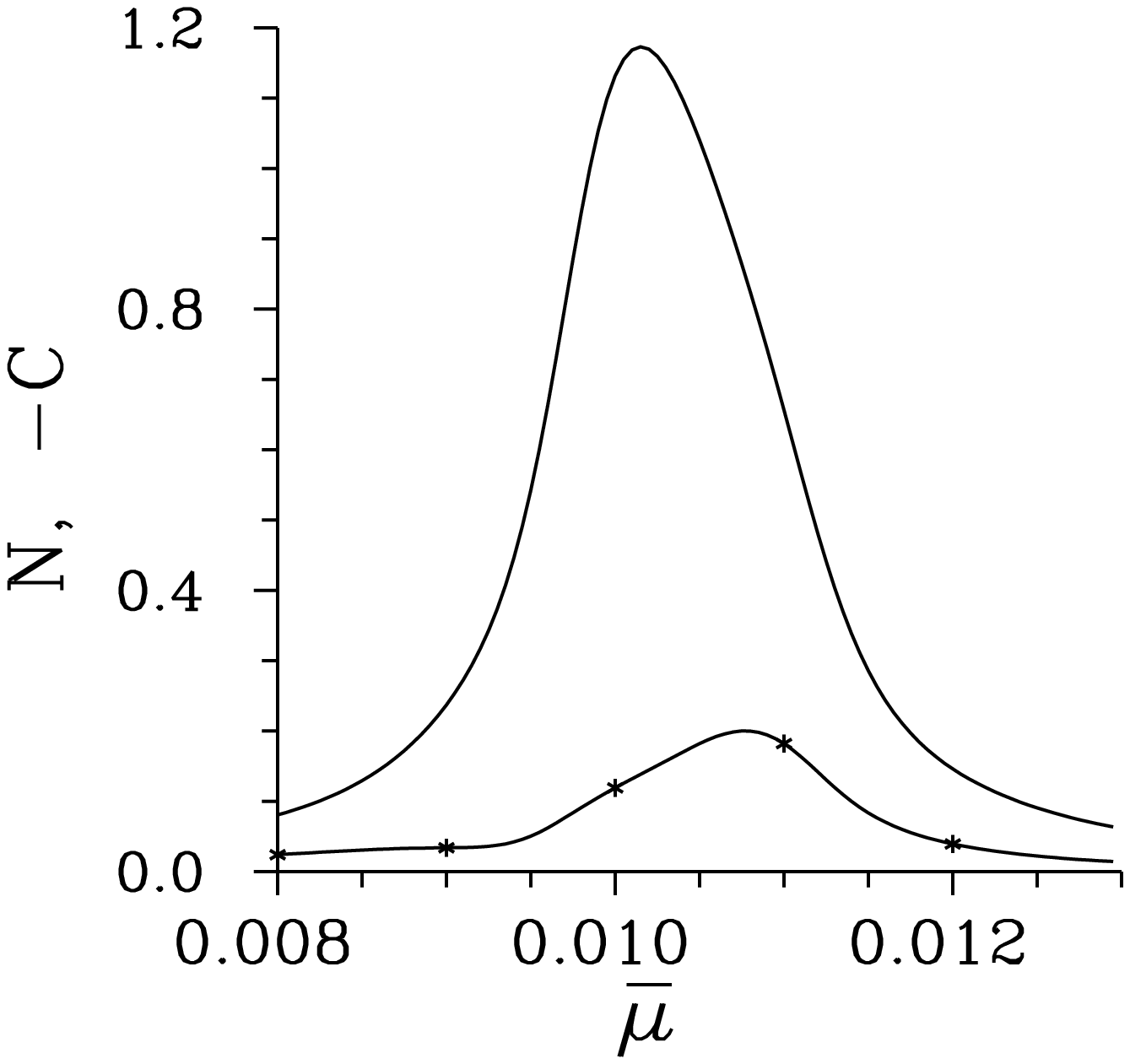}}

 \centerline{(a) \hspace{.4\hsize} (b)}
 \end{center}
 \caption{Quadratic negativity $ N $ (curve without symbols) and
  covariance $ C $ (curve with $ \ast $) as they depend on dipole
  moments $ \tilde{\mu} \equiv \tilde{\mu}_a = \tilde{\mu}_b $ for
  (a) $ J_a = J_b = 0.001 $,  $ J_{ab} = 0 $ and (b) $ J_{ab} = 0.001 $,
  $ J_a = J_b = 0 $;
  $ E_a = E_b = E_L = 1 $, $ \mu_a = \mu_b = 0.001 $, $ V_a = V_b
  = 0.01 $, $ \alpha_L = 1 $.}
\label{fig11}
\end{figure}

Molecular condensates \cite{Silinsh1994} represent a typical
example. In molecular crystals, energies of discrete excited
states are in 1~eV, Coulomb configuration interaction energies in
10~meV and dipole-dipole interaction energies in $ 0.1 \sim 1
$~meV. Typical values of dipole moments are expressed in 1~D.
Electric-field amplitudes in $ 1 \times 10^8 $~V/m are thus
necessary to arrive at comparable dipole and Coulomb configuration
interaction energies needed for the observation of the
Autler-Townes splitting. If the Coulomb configuration interaction
energy ($ V_a $) and the direct dipole interaction energy ($
\bar\mu_a \alpha_L $) have equal strengths, also the ionization
process considerably slows down. Electrons then tend to occupy
their ground states and discrete auto-ionizing states, i.e. states
mutually interacting through the dipole-dipole interaction.
Alternatively, the electrons can undergo ionization caused by the
dipole-dipole interaction with the continua. In both cases, the
influence of dipole-dipole interaction to the ionization process
dramatically increases. This results in greater values of
negativities $ N $ [see Fig.~\ref{fig11}(a) for the interaction
with the continua and Fig.~\ref{fig11}(b) for the interaction
between the auto-ionizing levels].

\begin{figure}        
 \centerline{\resizebox{0.7\hsize}{!}{\includegraphics{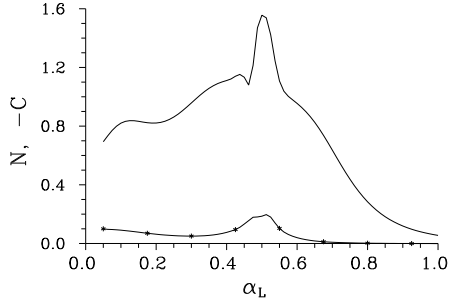}}}
 \caption{Quadratic negativity $ N $ (curve without symbols) and
  covariance $ C $ (curve with $ \ast $) depending on pump-field
  amplitude $ \alpha_L $;
  $ E_a = E_b = E_L = 1 $, $ \mu_a = \mu_b = 0.001 $, $ V_a = V_b
  = 0.01 $, $ \tilde{\mu}_a = \tilde{\mu}_b = 0.02 $, $ J_{ab} = 2 \times
  10^{-4} $, $ J_a = J_b = 0 $.}
\label{fig12}
\end{figure}

Considering fixed values of parameters $ V $ and $ \tilde\mu $,
the conditions maximizing negativity $ N $ and discussed above can
be reached choosing suitably the pump-field amplitude $ \alpha_L
$. This is possible because negativity $ N $ as a function of
pump-field amplitude $ \alpha_L $ exhibits a well-formed maximum
(see Fig.~\ref{fig12}).

\section{Conclusions}

Analytical solution of the model of two auto-ionization systems
interacting through the dipole-dipole interaction has been found
using the Laplace transform of the dynamical equations. Quadratic
negativity together with its spectral density have been defined to
quantify entanglement between two ionized electrons developed due
to the dipole-dipole interaction. The dipole-dipole interaction
with the continua weakens the interference of direct and indirect
ionization paths. This results in spectral shifts of the peaks in
the long-time photoelectron ionization spectra towards the energy
of the auto-ionizing state. On the other hand, the dipole-dipole
interaction between the auto-ionizing states leads to the
occurrence of an additional peak in the long-time photoelectron
ionization spectra centered at the energy of the auto-ionizing
state. When the auto-ionizing states play a dominant role in
ionization and a strong pump field induces the Autler-Townes
splitting of the ionization peak, the dipole-dipole interaction
with the continua weakens this splitting. Balance of the
dipole-dipole interactions with the continua and between the
auto-ionizing states leads to the occurrence of a minimum in the
long-time photoelectron ionization spectrum of an individual atom
that resembles the Fano zero. The dipole-dipole interactions are
responsible for strong anti-correlations in energies of the
ionized electrons. They originate in the energy conservation. Both
dipole-dipole interactions participate together in the creation of
entanglement. The stronger the dipole-dipole interactions are the
more entangled the state is. Optimal conditions for entanglement
creation occur when the direct and indirect ionization paths are
balanced and slow ionization is observed. 'Distribution of
entanglement' in the state of two ionized electrons as described
by the density of quadratic negativity reflects the shape of the
corresponding joint photoelectron ionization spectrum.

\ack

J.P. and A.L. gratefully acknowledge the support by the project
LO1305 of the Ministry of Education, Youth and Sports of the Czech
Republic. J.P. and W.L. also thank Operational Program Education
for Competitiveness - European Social Fund project
CZ.1.07/2.3.00/20.0058 of the Ministry of Education, Youth and
Sports of the Czech Republic. A.L. thanks the project IGA No.
PrF/2014/005. W.L. acknowledges Vietnam Ministry of Education
Grant No. B2014-42-29 for support.

\appendix

\section{Solution for two independent auto-ionization systems}

If atoms $ a $ and $ b $ are independent the quantum state $
|\psi\rangle(t) $ in Eq.~(\ref{5}) can be written as
\begin{equation}   
 |\psi\rangle^{\rm ind}(t) = |\psi_a\rangle(t) |\psi_b\rangle(t).
\label{A1}
\end{equation}
In Eq.~(\ref{A1}) states $ |\psi_j\rangle(t) $, $ j=a,b $,
describe electrons at individual atoms:
\begin{eqnarray}  
 |\psi_j\rangle(t) = c^j_{0}(t) |0\rangle_{j} + c^j_{1}(t) |1\rangle_{j}
  + \int dE_j d_j(E_j,t) |E_j\rangle. \nonumber \\
 \label{A2}
\end{eqnarray}

The corresponding differential equations for the coefficients in
Eq.~(\ref{A2}) are derived as follows:
\begin{eqnarray} 
 i \frac{d}{dt} \left[ \begin{array}{c} {\bf c}_j(t) \cr d_j(E_j,t)
  \end{array} \right] = \left[ \begin{array}{cc}
  {\bf K}_j & \int dE_j {\bf I}_j \cr
  {\bf I}_j^\dagger & E_j-E_L \end{array} \right]
  \left[ \begin{array}{c} {\bf c}_j(t) \cr d_j(E_j,t)
  \end{array} \right] \nonumber \\
\label{A3}
\end{eqnarray}
using vectors $ {\bf c}_j $, $ {\bf c}_j^T = (c^j_0,c^j_1) $ for $
j=a,b $, and matrices $ {\bf K}_j $ and $ {\bf I}_j $ defined in
Eq.~(\ref{7}).

The Laplace transform leaves equations~(\ref{A3}) in the form:
\begin{eqnarray} 
 \left[ \varepsilon {\bf 1} - {\bf K}_j \right] \tilde{\bf c}_j(\varepsilon)
  - \int dE_j \, {\bf I}_j \tilde{d}_j(E_j,\varepsilon) = i c_j(0) , \nonumber \\
 \left(\varepsilon - E_j+E_L\right) \tilde{d}_j(E_j,\varepsilon)
  - {\bf I}_j^\dagger \tilde{\bf c}_j(\varepsilon) = 0 ,  \hspace{5mm} j=a,b. \nonumber \\
\label{A4}
\end{eqnarray}
Applying the approach similar to that used in Sec.~2 the solution
of Eqs.~(\ref{A4}) is revealed in the form:
\begin{eqnarray} 
 \tilde{\bf c}_j(\varepsilon) = i \sum_{k=1,2}
  \frac{ {\bf L}^j_k {\bf c}_j(0)}{ \varepsilon + \lambda^j_k}  ,
  \nonumber \\
 \tilde{d}_j(E_j,\varepsilon) = \frac{i}{\varepsilon - E_j+E_L} \sum_{k=1,2}
  \frac{ {\bf I}_j^\dagger {\bf L}^j_k {\bf c}_j(0)}{ \varepsilon +
  \lambda^j_k}, \hspace{5mm} j=a,b.  \nonumber \\
\label{A5}
\end{eqnarray}
Matrices $ {\bf L}^j_k $ and eigenvalues $ \lambda^j_k $ used in
Eq.~(\ref{A5}) are given in Eqs.~(\ref{19}) and (\ref{20}),
respectively. The inverse Laplace transform provides this solution
in the time domain:
\begin{eqnarray} 
 {\bf c}_j(t) = \sum_{k=1,2}
   {\bf L}^j_k {\bf c}_j(0) \exp(i\lambda^j_k t) , \nonumber\\
  d_j(E_j,t) = \sum_{k=1,2}
  \frac{ {\bf I}_j^\dagger {\bf L}^j_k {\bf c}_j(0)}{ E_j-E_L + \lambda^j_k}
   \Bigl\{ \exp[-i(E_a-E_L)t] \nonumber  \\
 \mbox{} \hspace{10mm} - \exp(i\lambda^j_k t) \Bigr\} , \hspace{5mm} j=a,b.
\label{A6}
\end{eqnarray}

Considering the common state $ |\psi\rangle^{\rm ind} $ of both
atoms defined in Eq.~({A1}), the Laplace transform of coefficients
$ {\bf c}(t) $ defined by direct product $ {\bf c}_a(t) \otimes
{\bf c}_b(t) $ is expressed as:
\begin{equation} 
 \tilde{\bf c}(\varepsilon) = i \sum_{k,k'=1,2}
  \frac{[{\bf L}^a_k {\bf c}_a(0)]\, [{\bf L}^b_{k'} {\bf c}_b(0)]}{
  \varepsilon + \lambda^a_k + \lambda^b_{k'}} .
\label{A7}
\end{equation}
Similarly, the Laplace transform of coefficients $ {\bf
d}_a(E_a,t) $ determined by direct product $ {\bf c}_b(t) \otimes
d_a(E_a,t) $ takes the form:
\begin{eqnarray} 
 \tilde{\bf d}_a(E_a,\varepsilon) = i \sum_{k,k'=1,2}
  \frac{[{\bf I}_a^\dagger {\bf L}^a_k {\bf c}_a(0)]\, [{\bf L}^b_{k'} {\bf c}_b(0)]}{
  E_a-E_L + \lambda^a_k } \nonumber \\
 \hspace{10mm} \mbox{} \times \left[ \frac{1}{\varepsilon -E_a+E_L + \lambda^b_{k'}} -
  \frac{1}{\varepsilon + \lambda^a_k + \lambda^b_{k'} } \right] .
\label{A8}
\end{eqnarray}

Using solutions (\ref{A7}) and (\ref{A8}), the first term in
Eq.~(\ref{13}) can be treated as follows:
\begin{eqnarray}  
 \left[ (\varepsilon - E_a + E_L) {\bf 1} + {\bf L}_b \right]
  \tilde{\bf d}_a(E_a,\varepsilon) = \nonumber \\
 \sum_{k=1,2}(\varepsilon - E_a + E_L + \lambda_k^b){\bf L}^b_k
  \tilde{\bf d}_a(E_a,\varepsilon) = \nonumber \\
 i \sum_{k,k'=1,2} \frac{[{\bf I}_a^\dagger {\bf L}^a_k {\bf c}_a(0)]
  [{\bf L}^b_{k'} {\bf c}_b(0)]}{ \varepsilon + \lambda^a_k +
  \lambda^b_{k'}} = {\bf B}_a^\dagger \tilde{\bf c}(\varepsilon) .
 \hspace{5mm}
\label{A9}
\end{eqnarray}
The following relations have been used in Eq.~(\ref{A9}),
\begin{equation} 
 {\bf L}^a_k {\bf L}^a_{k'} = \delta_{kk'} {\bf L}^a_k ,
  \hspace{10mm} {\bf B}_a = \left[ \begin{array}{cc} {\bf I}_a & {\bf 0} \cr
  {\bf 0} & {\bf I}_a \end{array} \right] .
\label{A10}
\end{equation}
The equality expressed in Eq.~(\ref{A9}) implies that the integral
term in Eq.~(\ref{13}) is zero when independent atoms $ a $ and $
b $ are considered. The same holds also for the integral term in
Eq.~(\ref{14}) owing to the symmetry $ a \leftrightarrow b $. We
note that this conclusion can be achieved also by direct
integration of the solution written in Eq.~(\ref{A8}).

\section{Competition of the discrete and continuum dipole-dipole interactions}

Spectra in auto-ionization systems are formed by mutual
interference of two types of ionization paths. One of them is
based upon direct ionization originating from the ground state.
The other uses an excited auto-ionizing discrete level that
mediates auto-ionization of the system. These two ionization paths
may interfere destructively. The presence of the Fano zero in the
ionization spectra \cite{Fano1961} represents a clear evidence of
such quantum interference. Similarly, the dipole-dipole
interaction of an auto-ionization system with its neighbor can be
divided into two parts \cite{PerinaJr2011a}. One part influences
the discrete auto-ionizing level, states of the continuum are
modified by the other. Also these two parts of the interaction
compete. However, this competition occurs at the level of
population of the involved quantum states as the dipole-dipole
interaction means energy transfer. This competition naturally
modifies photoelectron spectra. The greatest changes in
photoelectron spectra are observed when both parts of the
dipole-dipole interaction have comparable strengths. Suitable
conditions for this case can be revealed as follows. First, we
apply the usual Fano unitary transformations of excited/ionized
states at atoms $ a $ and $ b $. The appropriate values of
parameters are then found from the requirement that the
interaction Hamiltonian $ \hat{H}_{\rm trans} $ given in
Eq.~(\ref{4}) is zero. In this case, two parts of the
dipole-dipole interaction related to the discrete and continuum
states of both atoms are equally strong but with the opposed
signs.

Invoking the Fano unitary transformations
\cite{Fano1961,PerinaJr2011b} on both atoms, Hamiltonians $
\hat{H}_j^0 $ written in Eq.~(\ref{2}) are transformed into their
diagonal forms,
\begin{equation}  
 \hat{H}_j^0 = \int dE_j E_j |E_j)(E_j| ,
  \hspace{5mm} j=a,b.
\label{B1}
\end{equation}
In Eq.~(\ref{B1}), states $ |E_j) $ arise from the Fano
diagonalization,
\begin{eqnarray}   
 |E_j) = f_j(E_j) |1\rangle_j + \int dE'_j \,
  g_j(E_j,E'_j)|E_j'\rangle , \hspace{5mm} j=a,b
  \nonumber \\
\label{B2}
\end{eqnarray}
and
\begin{eqnarray}   
 f_j(E_j) = \frac{ V^*_j(E_j) }{E_j-\tilde{E}_j^0+i\gamma_j},
  \nonumber \\
 g_j(E_j,E'_j) = \frac{V_j(E'_j)f_j(E_j)}{E_j-E'_j+i\epsilon } +
  \delta(E_j-E'_j) .
\label{B3}
\end{eqnarray}
Damping constants $ \gamma_j $, $ \gamma_j = \pi |V_j|^2 $, and
shifted energies $ \tilde{E}_j^0 $, $ \tilde{E}_j^0 = E_j^0 +
{\cal P} \int dE |V_j(E)|^2 / (E_j^0-E) $, have been introduced in
Eq.~(\ref{B3}). Symbol $ \epsilon $ denotes a small positive
number and limit $ \epsilon \rightarrow 0 $ is assumed at the end
of calculations.

After the transformations, Hamiltonians $ H_j^L $ defined in
Eq.~(\ref{3}) attain the form:
\begin{equation}   
 \hat{H}_j^L = \int dE_j \bar{\mu}_j(E_j) \alpha_L\exp(-iE_L t) |E_j)\,{}_j\langle
 0|+\mbox{H.c.}
\label{B4}
\end{equation}
Dipole moments $ \bar\mu_j(E_j) $ describing excitation/ionization
of states $ |E_j) $ inside the structured continua are derived in
the form
\begin{equation}   
 \bar{\mu}_j(E_j) = \tilde\mu_j \frac{ \epsilon_j(E_j) + q_j }{\epsilon_j(E_j)
 - i};
\label{B5}
\end{equation}
$ \epsilon_j(E) = (E -\tilde{E}_j^0)/\gamma_j $, $ q_j = \mu_j
/(\pi\tilde\mu_j V_j^*) $, and $ i $ stands for the imaginary
unit. If the dipole-dipole interaction between two atoms is
neglected, we have $ \bar\mu_a(E_a^F) = \bar\mu_b(E_b^F) = 0 $ for
$ E_a^F - \tilde{E}_a^0 = - \gamma_a q_a $ and $ E_b^F -
\tilde{E}_b^0 = - \gamma_b q_b $. Thus, there occurs one Fano zero
with energy $ E_a^F $ in the photoelectron ionization spectrum of
atom $ a $ and one Fano zero with energy $ E_b^F $ in the spectrum
of atom $ b $.

These Fano zeros are concealed by the dipole-dipole interaction.
The greatest competition of two parts of the dipole-dipole
interaction is observed provided that the matrix element of
transfer Hamiltonian $ \hat{H}_{\rm trans} $ between states $
|E_a^F) $ and $ |E_b^F) $ equals zero. Hamiltonian $ \hat{H}_{\rm
trans} $ given in Eq.~(\ref{4}) is expressed in these bases as
\begin{equation}   
 \hat{H}_{\rm trans} = \int dE_a \int dE_b \bar{J}(E_a,E_b)
  |E_b) (E_a| + {\rm H.c.},
\label{B6}
\end{equation}
where
\begin{eqnarray}   
 \bar{J}(E_a,E_b) = J_{ab} f_a(E_a) f_b^*(E_b) \nonumber \\
 \hspace{10mm} \mbox{} + \int dE'_a \,J_a^* g_a(E_a,E'_a) f_b^*(E_b) \nonumber \\
 \hspace{10mm} \mbox{} + \int dE'_b \,J_b f_a(E_a) g_b^*(E_b,E'_b) .
\label{B7}
\end{eqnarray}
The use of expressions written in Eqs.~(\ref{B3}) transforms
equation~(\ref{B7}) into the form:
\begin{eqnarray}  
 \bar{J}(E_a,E_b) = \frac{J_a^* V_b}{\gamma_b[\epsilon_b(E_b)-i]}
   + \frac{J_b V_a^*}{\gamma_a[\epsilon_a(E_a)+i]} \nonumber \\
 \hspace{10mm} \mbox{} + \frac{J_{ab}V_a^*V_b - iJ_a^* V_b\gamma_a + iJ_b V_a^*\gamma_b}{
  \gamma_a\gamma_b[\epsilon_a(E_a)+i][\epsilon_b(E_b)-i]} .
\label{B8}
\end{eqnarray}

The requirement $ \bar{J}(E_a^F,E_b^F) = 0 $ results in the
following condition for the values of dipole-dipole interaction
constants $ J_{ab} $, $ J_a $ and $ J_b $:
\begin{equation}  
 \frac{J_a^* \mu_a^*}{ \tilde\mu_a^*} + \frac{J_b \mu_b}{ \tilde\mu_b}
  = J_{ab} .
\label{B9}
\end{equation}
Assuming identical atoms $ a $ and $ b $ and real interaction
constants, the condition in Eq.~(\ref{B9}) simplifies
\begin{equation}  
 \frac{J_a}{ J_{ab}/2} = \frac{ \tilde\mu_a}{\mu_a}.
\label{B10}
\end{equation}

\section*{References}


\end{document}